\documentclass[pra,superscriptaddress]{revtex4}

\newcommand{\epigram}[3]
{
\begin{flushright}
\baselineskip=13pt
\parbox{4.2in}{\baselineskip=13pt
``#1''}\medskip\\
---{\it #2}\\
#3
\end{flushright}
}

\usepackage{natbib}

\usepackage{amssymb}
\usepackage{amsfonts}
\newcommand{\text}{\mbox}

\def\Q{{\mathbb{Q}}}
\def\H{{\mathbb{H}}}
\def\R{{\mathbb{R}}}
\def\N{{\mathbb{N}}}
\def\C{{\mathbb{C}}}

\def\OR{\vee}

\def\union{\cup}

\def\fsl{\mathfrak{sl}}
\def\fsu{\mathfrak{su}}

\def\openone{\leavevmode\hbox{\small1\kern-3.8pt\normalsize1}}
\def\RR{{\rm I\kern-.2emR}}
\def\tr{{\rm tr}\; }
\def\ce{{\mathcal E}}

\def\cd{{\mathcal D}}
\def\cb{{\mathcal B}}

\def\ch{{\mathcal H}}
\def\cs{{\mathcal S}}
\def\ct{{\mathcal T}}
\def\cp{{\mathcal P}}
\def\cm{{\mathcal M}}
\def\cf{{\mathcal F}}
\def\ci{{\mathcal I}}
\def\ca{{\mathcal A}}
\def\co{{\mathcal O}}
\def\calr{{\mathcal R}}
\def\cv{{\mathcal V}}
\def\cz{{\mathcal Z}}

\newcommand{\ket}[1]{| #1 \rangle}
\newcommand{\bra}[1]{\langle #1 |}

\newcommand{\outerp}[2]{\ket{#1}\! \bra{#2}}

\newcommand{\projj}[2]{\ket{#1}\ket{#2}\! \bra{#2}\bra{#1}}
\newcommand{\kett}[2]{\ket{#1}\ket{#2}}
\newcommand{\braa}[2]{\bra{#1}\bra{#2}}

\newcommand{\beq}{\begin{equation}}
\newcommand{\eeq}{\end{equation}}
\newcommand{\beqa}{\begin{eqnarray}}
\newcommand{\eeqa}{\end{eqnarray}}
\newtheorem{definition}{Definition}
\newtheorem{problem}{Problem}
\newtheorem{theorem}{Theorem}
\newtheorem{proposition}{Proposition}

\newtheorem{conjecture}{Conjecture}

\newtheorem{axiom}{Axiom}

\def\QED{\mbox{\rule[0pt]{1.5ex}{1.5ex}}}

\begin{document}
\raggedbottom

\title{COORDINATING QUANTUM AGENTS' PERSPECTIVES:
CONVEX OPERATIONAL THEORIES, QUANTUM INFORMATION, 
AND QUANTUM FOUNDATIONS}

\author{H. BARNUM}
\address{Computer and Computational Sciences Division CCS-3, MS B256,\\
  Los Alamos National Laboratory, Los Alamos 87545, USA\\
  E-mail: barnum@lanl.gov}

\begin{abstract}

In this paper, I propose a project of enlisting quantum information
science as a source of task-oriented axioms for use in the
investigation of operational theories in a general framework capable
of encompassing quantum mechanics, classical theory, and more.
Whatever else they may be, quantum states of systems are
compendia of probabilities for the outcomes of possible
operations we may perform on the systems: ``operational theories.''  I
discuss appropriate general frameworks for such theories, in
which convexity plays a key role.  Such frameworks are appropriate
for investigating what things look like from an ``inside
view,'' {\em i.e.} for describing perspectival information that one
subsystem of the world can have about another.  Understanding how such
views can combine, and whether an overall ``geometric'' picture
(``outside view'') coordinating them all can be had, even if this
picture is very different in nature from the structure of the
perspectives within it, is the key to understanding whether we may be
able to achieve a unified, ``objective'' physical view in which
quantum mechanics is the appropriate description for certain
perspectives, or whether quantum mechanics is truly telling us we must
go beyond this ``geometric'' conception of physics.  The nature of  
information, its flow and processing, as seen from various operational
persepectives, is likely to be key to understanding whether and how
such coordination and unification can be achieved.

\end{abstract}

\maketitle

\epigram{The subject...had in fact become an obsession and, like all 
obsessions, very likely a dead end.  It was the kind of favorite puzzle that
keeps forcing its way back because its very intractability makes it
perversely pleasant.... I hoped that some new image might propel me past
the jaded puzzle to the other side, to ideas strange and compelling.}
{E.O. Wilson}{The Diversity of Life}

\section{Introduction}

The central question quantum mechanics raises for the foundations of
physics is whether the attempt to get a {\em physical} picture, from
``outside'' the observer, of the observer's interaction with the
world, a picture which views the observer as part of a reality which
is at least roughly described by some mathematical structure which is
interpreted by pointing out where in this structure we, the observers
and experimenters, show up, and why things end up looking as they do
to observers in our position, is doomed.  The ``relative state''
picture that arises when one tries to describe the whole shebang by an
objectively existing quantum state is unattractive, and many seek to
interpret quantum states instead as subjective, ``information'' about
how our manipulations of the world could turn out.  Whatever else they
may be the quantum states of systems clearly {\em are} compendia of
probabilities for the outcomes of possible operations we may perform
on the systems: ``operational theories.''  Quantum information science
concerns what one can do with the operations available to us quantum
mechanically, and has shown the power of the formal analogy between
quantum states and classical information (probability distributions), 
an analogy which is also at the heart of the ``subjectivist'' attitude
\cite{CFS2002a} towards quantum states.
Quantum computation and information theory are making the movement
between ``inside'' and ``outside'' points of view of measurement 
processes and other interactions, formulated in a general and rigorous
way, a commonplace of physical practice.

In this paper, I will suggest a particular project for harnessing
quantum information-theoretic and computational ideas to the task of
resolving the central tension in the foundations of quantum mechanics.
This project has the virtue that even if it does not succeed in
resolving this tension, the technical results obtained in the attempt
will still be valuable.  The idea is to view quantum mechanics as a
framework for a particular type of
{\em empirical, operational} theories.  Such a theory 
is, roughly, a compendium of operations (such as measurements)
that can be performed on a system, together with a specification of
the probabilities of their outcomes, usually called a ``state.''
Actually, I will sometimes refer to as an {\em operational theory},
what should perhaps be called an ``operational theory-type'': a
specification of the possible set of operations on a system, together
with a specification of a set of {\em possible} compendia of
probabilities for results of these operations.  An ``operational'' or
``phenomenological'' theory itself arises when a particular
probability compendium---a particular state---is chosen.  
  Some who view quantum states as
subjective and have a generally Bayesian view of probability as well,
have expressed discomfort with the notion of starting with probabilities
as ``empirically given,'' but of course one can and should view such 
``phenomenological'' probabilities as themselves established via 
Bayesian inference.  It is true that, in the case of quantum mechanics,
for example, our actual subjective
probabilities about the outcomes of experiments might not coincide
precisely with the quantum-mechanical ones, because even after taking
all the evidence into account we might have some
very small subjective probability that quantum mechanics is wrong.
I don't think this causes any major difficulties for the approach
taken below, however.

The viewpoint just sketched is the 
``operational'' or ``operational quantum logic'' point
of view, and it can be useful whether one wants to
consider the empirical theory as for some reason all we can hope for, 
{\em or} as a description
of how perspectives look within an overarching theory such as the
relative state interpretation (RSI).  
Therefore, technical results obtained using this point of 
view are likely to be valuable to both camps, and of interest regardless
of the ultimate disposition, if any, of the problem of quantum 
foundations.  Nevertheless, they could also turn out to have implications
for the disposition of this problem.

The new light being shed by
quantum information science on foundations has, {\em so far}, served
primarily to put the old questions in brighter relief, to pose them
more clearly and in some cases to provide a framework, information
theory, in which the peculiar features of quantum mechanics can be
expressed quantitatively.  In a way, this just illuminates even more
clearly the stalemate reached early on by those attempting to
understand quantum foundations.  The fundamental issue is the tension
between quantum-mechanical descriptions of measurement processes from
the ``inside'' and ``outside'' points of view:  from the point of
view of the measuring system (observer) or from outside both observer
and observed.  Is the there an acceptable quantum-mechanical
outside
view, and if not,
must be content, perhaps even view ourselves as
lucky, to live without such an outside view?  Most commonly, this
takes the form of controversy between those espousing a ``relative
state'' interpretation of quantum mechanics
\cite{Everett57b, Everett57a}, 
in which the state vector
is interpreted as a real physical object, extendible to include
whatever observers may be involved in a measurement, and those
espousing an ``information'' view of quantum mechanics, in which the
statevector describes, roughly, the information about possessed by an
observer who is in a position to do experiments on a system, about the
possible outcomes of those experiments {\em and} should not be
interpreted as having any more objective nature than this.  While it
has in my view as yet not made a decisive contribution toward
resolving this tension, by focussing on the role of information held
(through entanglement or correlation) or obtained (by measurement) by
one system about another, it concentrates one's attention on the
practical importance of such measurements, and develops flexibility in
moving between the inside and outside views of such
information-gathering processes.  It thus provides tools and concepts,
as well as the ever-present awareness, likely to be useful in
resolving this tension, if that is possible.  In the best scenario, it
might help provide a ``new image'' of the sort hoped for in the epigram
from E.~O. Wilson that heads this section, to propel us, in his
somewhat Morrisonian words, ``past the jaded [{\it sic}] puzzle to the
other side, to ideas strange and compelling.''

Before discussing the project of combining the operational approach
with the new tools, images, and analogies provided by quantum information
science (QIS) in the
task of trying to coordinate quantum perspectives
into an overall physical picture, I consider, in Section \ref{sec:
peculiar}, some salient general implications of QIS for foundational
questions (irrespective of its contributions to this project).  QIS's
most impressive achievements are due precisely to the peculiarities of
quantum mechanics that have impressed quantum physicists since the
beginning.  While it may not have achieved any intepretational
breakthroughs, I illustrate how it provides tools for sharper analysis
of interpretational questions, raises some specific and somewhat
technical questions involving computability, and draws attention to
the fruitfulness of formal analogies between density matrices and
probability distributions.

In Sections \ref{sec: RSI vs subjective} and \ref{sec: laws of thought}
I expand upon the basic interpretational tension in the
foundations of quantum mechanics; this provides a necessary basis for
the discussion that follows, in general terms in Sec. \ref{sec:
combination of perspectives} and in more particulars in subsequent
sections, of how the marriage of QIP and empirical quantum logic might
contribute to its resolution.  (However, those whose only interest
is in technical aspects related to operational theories might skip
to Section \ref{sec: operational theories}.)

Section \ref{sec: operational theories} turns to what I
consider a good framework, centered around probabilities, for
empirical operational theories.  I use a notion of probabilistic
equivalence with a long history, that identifies outcomes of different
procedures (as far as ``their effect on the observer'' is concerned)
if they have the same probability in all states of the
phenomenological theory.  I show it allows one to derive, from a
phenomenological operational theory, a more abstract structure, a
``weak effect algebra,'' closely related to (and probably completable
to) a structure well-studied in operational quantum logic:  that of
``effect algebra.''  
 I review these and related notions of empirical
quantum logic.   In Section 7 I review 
```convex effect algebras,'' emphasizing how
the modern ``positive operator valued measures'' (POVMs) approach to quantum
mechanics (in terms of POVMS's and trace-nonincreasing completely positive
maps, aka
``quantum operations'') is a case of these, and how foundational
results involving it are naturally expressed in these terms, sometimes
as cases of general results involving convex effect algebras.
 I discuss
probabilistic equivalence in such a setting, and relate Gleason-type
theorems such as the recent ones for POVMs, to general results locating
the states in the dual of the
cone on which the convex effect algebra can be represented.

In Section \ref{sec: operation algebras}
 I sketch work in progress on how such a framework might be
extended to incorporate dynamics conditional on measurement results,
exhibiting the quantum mechanics not only of effects but also of
``operations,'' so
familiar now to physicists through its ubiquity in the theory of QIP,
as an example of a structure I call an {\em operation algebra},
basically a weakening (to have multiple top elements, corresponding to the
many possible trace-preserving operations) of the notion of effect
algebra, with the addition of a product of operations, 
interpreted as doing the operations in sequence, with the second
conditioned on the first.
In section \ref{sec: combination} I discuss the combination of 
subsystems in operational theories, and its relationship to 
dynamics in such theories and to the ``coordination of perspectives.''
In Section \ref{sec: dynamics} I discuss dynamics further, showing
that there is a potential difference between 
``Schr\"odinger/Liouville/vonNeumann'' dynamics that act on states,
and ``Heisenberg'' type ones that act on an effect algebra, arguing
that only those Schr\"odinger dynamics that have a Heisenberg 
representation should be countenanced, at least in theories whose
state space is maximal.
In Section \ref{sec: applying operational logic}, I discuss how this
framework may be used in the project of applying QIP ideas to
foundational questions.  This involves several thrusts.  One is to
develop, from a general operational theories approach, an
understanding of the various ways subsystems may be combined to form
larger systems, or individuated as parts of such systems.  If quantum
theory is viewed as a framework for ``perspectives'' of one subsystem
on another, such a theory of subsystem combination is essential to the
project of combining such perspectives in an overarching ``geometric''
picture.  The category-theoretic tensor product and similar 
(but ``non-minimal'') constructions are of particular
interest as one such way of combining subsystems, probably leading (if
applied thoroughgoingly to quantum mechanics) to a standard
``relative-state'' view as the overarching picture; but the
consideration of other possibilities is probably necessary if one is
to avoid the grotesque features of that ``many-worlds'' theory, and
interestingly, it is suggested by some other attempts to resolve
quantum puzzles via taking a ``global'' point of view on quantum
theory: Schulman's (\citeyear{Schulman97a}) symmetric boundary
conditions proposal, and the Rovelli-Smolin (\cite{LSmolin95a}) approach
to unifying quantum mechanics and gravity with a ``topological quantum
field theory'' type construction (and the related ``relational'' approach
to quantum mechanics \cite{Rovelli96a}.  It could also turn out that rather
than, or along with, non-tensor combination of objects in some
situations, the appropriate category could change, with a quantum
description (as might also be the case with the tensor product
combination law) only being appropriate to perspectives in certain
situations or limits.

Even if no such overarching picture turns out to be possible and 
our ``empirical operational theories'' of how things look from 
various perspectives are viewed as all we are going to get, the 
consideration of subsystem combination is important in the understanding
of empirical operational theories.  One may reasonably want to 
require that the operations we perform on a system 
should be allowed to use auxiliary systems as apparatus, and we should
be able to include those systems with the observed system, and 
describe our perspective on the system and apparatus in the same
general terms (e.g. quantum mechanics) as the system itself.  Within
standard quantum mechanics, certainly, this works.  

A major part of the project involving ``empirical operational
theories'' is, as it has been since long before quantum computation
came on the scene, ``deriving quantum mechanics.''  The hope is that
if this can be done with axioms whose physical, or better yet
information-theoretic or information-processing, meaning is clear,
then one will have a particularly nice kind of answer to the question
``Why quantum mechanics?''  This has also been a preoccupation of many
of the QI/QC researchers who delve into foundations (e.g.
\cite{Fuchs2001a, Hardy2001a, Hardy2001b, Barnum2001a,Simon2001a}).  I
believe that QI/QC will provide a useful new source of axioms, with
natural interpretations involving the possibility or impossibility of
various information-theoretic tasks.  But, one may ask, how might this
contribute to resolving the basic tension?  I think it is likely to
contribute to {\em whichever} mode of resolution turns out to be
right.  Within the ``geometric'' or ``objective overall picture''
resolution, one might obtain the answer: Why quantum mechanics?
``Because it's the sort of structure you'd expect for describing the
perspectives [or at least certain perspectives, of the sort beings
like us wind up with in situations considered in standard quantum
theory] that occur ``from the inside point of view'' within an
overarching picture of {\em this} [fill in the blank] sort.''  The
blank might be filled in with a fairly particular overarching physical
theory, or the quantum formalism might arise from quite general
features likely to be common to many more particular proposals.  A
very similar answer might arise from the more ``subjectivist'' point
of view on quantum states.  Why quantum mechanics?  ``Because it's the
sort of structure you'd expect for describing the perspectives [or at
least certain perspectives...] that occur ``from the inside point of
view'' within a reality of {\em this} [fill in the blank] sort, which
reality is however not completely describable in physical terms, so
that these perspectives are as good as physics ever gets. [Explanation
of why perhaps supplied here.]''

From this point of view, both those who hope for an overarching
physical picture, or think it likely to emerge, including relative
state-ers who think quantum mechanics already provides its general
outlines, and those who think such an overarching {\em physical}
picture unlikely to emerge, can nevertheless fruitfully pursue
essentially the same project.  This is the project of using axiomatic
arguments involving the notion of ``operational theory'' to derive
quantum mechanics, to understand, conceptually, how it differs from or
is similar to other conceivable possibilities for such theories, and
the extent to which the quantum structure does or does not follow from
elementary conceptual requirements on such theories (one way in which
it could be ``a law of thought'') or, in a more Kantian or perhaps
``anthropic'' way, from the possibility of rational beings like us (a
different way in which it could be ``a law of thought'').  Of course,
one's view on how the basic interpretational conflict will be resolved
might influence the axioms one chooses to investigate, with the
subjectivists perhaps more inclined to axioms stressing the formal
analogies between density matrices and probability distributions,
quantum ``collapse'' and Bayesian updating of probability
distributions (\cite{Fuchs2001a}).  (But since on the
``overarching physical picture with perspectives'' view the
probabilities are also tied to a ``subjective,'' perspectival element,
the Bayesian analogy might be quite natural on this picture too.)

It thus doesn't seem necessary to me to commit in advance to an
opinion on how the resolution will be achieved.  What does seem likely
is that however it is achieved, the close link between ``empirical
operational theories'' and perspectival information that one subsystem
of the world can have about another, suggests that in the project of
understanding quantum mechanics as a framework appropriate to such
perspectives, the concepts of quantum information theory and quantum
information processing will play a major role.

\section{QIP: The power of the peculiar}
\label{sec: peculiar}
Virtually all of the main aspects of quantum mechanics exploited in
quantum information processing protocols have been understood for
decades to be important peculiarities of quantum mechanics by those
concerned with foundations.  Often this understanding goes back,
sometimes in an incomplete form, to the founders of quantum mechanics
itself.  \cite{Schroedinger35a} coined the term entanglement, and
emphasized its importance and its relationship to the nonuniqueness
of the decomposition of mixed quantum states into pure states, 
after Einstein, Podolsky, and Rosen
(\citeyear{Einstein35a}) drew out its implications of nonlocality.
The
importance of the idea that obtaining information about a quantum
system necessarily disturbs it was emphasized by \cite{Heisenberg27a}
and \cite{Bohr28a} (though the Heisenberg light microscope proved to
be a misleading illustration in some respects, suggesting incorrectly
a need for exchange of energy/momentum as the source of the
disturbance). \cite{Bohr28a} emphasized complementarity.  The
nonlocal correlations allowed by entanglement can be viewed as
precisely what is exploited by certain better-than-classical
communication complexity protocols \citep{Buhrman97a}; the necessity
of disturbance when information is gathered on a genuinely quantum
ensemble \citep{Fuchs95b, Barnum98d, Barnum2001a, Banaszek2001a,
Bennett94a, Barnum2001b}, which is closely related to the ``no-cloning
theorem'' \citep{Wootters82a} and no-broadcasting
theorem \citep{Barnum96a, Lindblad99a}, is the basis of quantum cryptography;
the ability to obtain information complementary to that available in
the standard computational basis, by effectively (and efficiently!) 
measuring in a
Fourier-transformed basis to obtain information about the period of a
function, is the heart of the historic series of algorithms due to
\citet{Deutsch85a}, \cite*{Deutsch92a},
Bernstein and Vazirani's (\citeyear{Bernstein97a})
algorithm  for ``recursive Fourier sampling,'' (the
first superpolynomial quantum query speedup in the bounded-error
model),
\citet{Simon97a}, and culminating in Shor's (\citeyear{Shor94a,Shor97a})
polynomial-time factoring algorithm.  
, also uses complementarity via the quantum Fourier transform.
The significance of these peculiarities of quantum mechanics is
brought out by quantum information science: these are no longer
curiosities, paradoxes, philosophers' conundrums, they now have
worldly power: to break public-key cryptography, speed up (though not
to polynomial time) brute-force search for solutions to, say,
NP-complete problems; to give unconditionally secure verification that
one's communication has not been overheard.  At the same time, the
concepts are made quantitative, analyzed more precisely.  Sometimes
their essential nature is somewhat better understood, but in my view
this has not yet led to breakthroughs in dealing with the
interpretational issues they raise (which is not to say such breakthroughs
may not come).

A number of more specific and/or technical points on which QIP has 
contributed, or shows potential to contribute, something
new to old debates can be identified.  Some of the more salient
ones are:

\begin{enumerate}
\item QIP provides tools with which to analyze much more precisely and
algorithmically questions of what can and cannot be measured, or
otherwised accomplished, either precisely or approximately, in quantum
mechanics.  Foundational work sometimes tends to take a rather
abstract view of quantum mechanics: systems correspond to Hilbert
spaces, states to density operators $\rho$ acting on those Hilbert
spaces, measurements to Hermitian operators $A$ (or decompositions of
unity) on those Hilbert spaces, probabilities of measurement outcomes
are given by $\tr \rho E_i,$ where $E_i$ is a projector onto the
subspace with eigenvalue $\alpha_i$ in the spectral decomposition $A =
\sum_i \alpha_i E_i$ of $A$ into orthogonal projectors (or by a positive
operator $E_i$ in a resolution of unity satisfying $\sum_i E_i = I$).
The question ``can any Hermitian operator be measured?'' has been a
longstanding one in foundations.  The Wigner-Araki-Yanase
theorem \citep{Wigner52a,Araki60a} sheds some light on it (a necessary
condtion for a Hermitian operator to be measured {\em precisely} is
that it commute with the system's free Hamiltonian), and also sheds
some light on the resources (notably interaction strength) needed to
approximately measure operators that can't be measured precisely.  
\citep*{Reck94a}
established that any operator could be measured using passive linear
optics (assuming arbitrary real parameters describing idealized
passive components can be achieved, of course an idealization but a
useful one).  Their scheme required exponential resources in the
number of qubits (or log of Hilbert space dimension) 
of the system to be measured.  The theory of universal quantum
computation shows, for the first time, that arbitrary operators can
be approximated arbitrarily well (and efficiently in the desired
precision), using a basis of local gates of size polynomial in the {\em
log} of the dimension of the Hilbert space (i.e., the number of
qubits).  (To forestall confusion: the claim is {\em not} that any
unitary can be approximated using a circuit containing a polynomial
number of gates; rather, the basis set of gates used can be taken to
be polynomial (say, a set of $g_1$ one-qubit and $g_2$ two-qubit gates,
applied on all qubits or pairs of qubits, for a basis of size $g_1 n +
g_2 n^2$ for expressing operators on $n$ qubits.))  Another point that
becomes clear when a computational approach is taken is that, given a
computational (quantum algorithmic) model of what a measurement is,
there will be Hermitian operators that are impossible to measure in
the same sense that certain functions are classically impossible to
compute (cf. \cite{Nielsen97a}, for example).  Indeed, prime examples in infinite-dimensional Hilbert
spaces are operators whose eigenvalues encode the values of an
uncomputable function, in the ``standard'' computational basis.  In
finite dimensional Hilbert spaces, the issue is less clear, but there
may well be operators that are unmeasurable essentially because some
of their eigenvalues or eigenvectors are uncomputable numbers (these
would probably have to be numbers that are not even computably
approximable, like Chaitin's $\Omega$).  Exactly what all this means
for the foundations of the theory, which are often taken to include
the abstract (and false) statement that all Hermitian operators can be
measured, is not clear, but there can be no doubt that it is
important.  It raises the issue of the extent to which ``operational''
limitations, including basic and highly theoretical ones such as
computability, should be built into our basic formalisms, and what it
means for the interpretation of those formalisms and the ``reality''
of the objects they refer to, if they are not.

\item QIP techniques and concepts promise to allow a much more systematic
approach than previously 
to experiments and thought-experiments suggested by foundational
investigations.  Error-correction, active and passive stabilization and
control techniques may allow
one to construct and quantum-coherently 
control much larger-dimensional quantum systems 
than heretofore, so that proposals that quantum coherence
is lost when systems become too large or too complex,
or other theories that postulate collapse as a real physical process
occuring in a specified manner under specified conditions, 
may be tested.  

Indeed, some (I am not among them) may suspect that the great 
difficulty so far experienced in getting even a few qubits of clean 
quantum computing power is a manifestation of such fundamental difficulties
with large-scale quantum coherence.  
Alternatively, we may come to see, if there arise 
insurmountable practical obstacles to implementing QIP, which are 
in turn understood as involving some general theoretical principles,
that macroscopic quantum coherence is in some sense a ``non-operational''
concept.  Whether that has repercussions for the interpretation of 
quantum mechanics (do both branches of Schr\"odinger's cat ``really 
exist'' in some sense) depends on how operational one is inclined to 
require one's ontology to be.  In my case, not particularly so, i.e. 
I am prepared to countenance the existence of theoretical objects whose
existence might be completely impractical, for theoretically clear reasons,
to operationally verify.  However, if one can get an adequate picture
of things without them, so much the better.  

\item QIP, especially quantum information theory (that branch of the
theory of QIP which is an analogue of classical information theory,
i.e. concerned particularly with asymptotic rates at which tasks such
as data compression and transmission can be performed, either
perfectly or as a function of a tolerable distortion level) has
demonstrated the power of taking the formal analogy between quantum
density matrices and classical probability distributions seriously.
Many things one does with probability distributions in classical
information theory (e.g. transmit messages from a source which emits
the messages with a given probability distribution), have natural
quantum analogues when quantum states replace probability
distributions.  This may have implications for foundations, as it goes
nicely with the view that ``the quantum state is essentially
information, and therefore collapse is not a mysterious, seemingly
causality-violating physical process, because it is not a physical
process, but a subjective one involving information.''  However, I
will argue below that this view of collapse is essentially shared by
the two main candidates for ``interpretations'' of nonrelativistic
quantum mechanics (the relative state interpretation and the
``no-interpretation'' interpretation).  Still, QIP thereby focusses
attention on these two approaches, which I view as corresponding to
``inside'' versus ``outside'' views of quantum measurement processes,
thereby throwing into sharper relief the fundamental issue in quantum
foundations, which is whether the outside view suggested by quantum
mechanics is really acceptable, and if not, whether it is acceptable
to live without such an outside view.

\item There is a long-standing philosophical/mathematical project of 
trying to understand quantum mechanics, and especially how it differs
from classical mechanics, in an abstract and axiomatic way. 
A goal of such an approach may be to find mathematically
natural characterizations of quantum theory and classical theory,
from within a framework of ``empirical theories'' more general than
either one.
Another goal is to find such characterizations that have clear 
physical interpretations.  QIP can contribute to this project by
providing a source of natural ``operational'' questions about
whether certain information-processing tasks can or cannot be
performed within the model theory, which may well be related to
natural axiomatic properties of the theory.  Also, QIP may be a natural
source of examples of such empirical theories.  These naturally
arise when one considers attempts to perform quantum information 
processing with restricted means, which may correspond to those
operations that are easy to do within a certain proposed implementation
of quantum computing (say, linear optics operations on a multi-mode
bosonic Hilbert space, with classical feedback \cite{KLM2001}), or to 
a theoretically natural restriction on 
quantum operations (say, those needed to encode, correct errors in,
and decode stabilizer quantum error correcting codes).  
For example, QIP considerations stimulated some of us \citep*{Barnum2002a}
to generalize
the notion of ``entanglement'' to a lie-algebraic situation more
general than that of standard multipartite quantum entanglement, 
and then to a ``convex cones'' setting reminiscent in some respects
of convex sets approaches to quantum foundations \citep{Davies70a,
Gudder99a,Ludwig81a,Ludwig81b}.
\end{enumerate}

\iffalse
[*maybe info disturbance made more quantitative (but probably
already essential facts realized, esp. in terms of arguments
surrounding heisenberg microscope...
*info as the appropriate language for formulating, making 
abstract and general, existing insights from foundations.
At the same time, making their power/applications evident.
*relevance to args about decoherence and biology.]
\fi

In the remainder of this paper, I will consider in detail this
last way in which QIP affects the study of foundations.  Inevitably,
though, it interacts with the other three points.  I will be
especially concerned with the relation to the
the inside/outside dichotomy, and relative-state versus the
``subjective density matrix'' interpretations, which I will 
relate to a tension between a ``geometric'' and ``dynamic'' view
of physics discussed, for example, in \cite{Bilodeau96a}. 
(I learned of this paper from a recommendation by Chris Fuchs.) 
For it is here, 
I think, that one finds suggestions of what else one needs from
physics to resolve the tension between these points of view,
and ideas for a research program aiming toward that goal.

In order to help understand how this goal might be realized I will
first present my take on the basic tension between ``inside'' and
``outside'' views of quantum mechanics, in the next section.

\section{In which Doctor Science meets Subjectiveman in his laser lab,
and turns quantum mechanics inside out}
\label{sec: RSI vs subjective}

\begin{flushright}
\baselineskip=13pt
\parbox{4.5in}{\baselineskip=13pt
``Day or night he'll be there anytime at all (Doctor Robert)\\
You're a new and better man \\
He helps you to understand \\
He does everything he can (Doctor Robert)\\
He's a man you must believe\\
No-one can succeed like Doctor Robert.''}\medskip\\
---{\it The Beatles}\\
Doctor Robert (Revolver)
\end{flushright}

\epigram{Don't pay money just to see yourself with Dr. Robert}{The Beatles}{Doctor Robert (Revolver)}
\epigram{
%Nor is this life of yours by which you set such store your
%doing, however you may choose to tell it.  Its shape was forced in the
%void at the onset and all talk of what might otherwise have been is
%senseless for there is no otherwise.  Of what could it be made?  Where
%be hid?  Or how make its appearance?  
The probability of the actual
is absolute.  That we have no power to guess it out beforehand 
makes it no less certain.
That we may imagine alternate histories means nothing at all.}
{Cormac McCarthy}{Cities of the Plain}

\iffalse
[Somewhere
in discussion of Bilodeau:  B's solution to S cat, (historical);
slippery slope problem;  history as hidden variables? Also: again
as related to slippery slope:  the Q phenomena don't take place in
Hilbert space, they take place in a lab.]
\fi

The central tension in interpreting quantum mechanics is between the
idea that we are part of a quantum world, made of quantum stuff
interacting with quantum stuff, evolving according to the
Schr{\"o}dinger equation, and the apparent fact that when we evolve so
as to correlate our state with that of some other quantum system
which is initially in a superposition, we get a single measurement
outcome, with probabilities given by the squared moduli of coefficients
of the projections of the state onto subspaces in which we see a 
definite measurement outcome.  The relative state interpretation attempts to
reconcile these ideas by taking the view that the experience of
obtaining a definite measurement result is simply how things appear
from one point of view, our subspace of the world's Hilbert space, and
the full state of the world is indeed a superposition.  As I see it
the correct way, on this view, to account for the appearance that
there is a single measurement result, is the idea that the experience
of a conscious history is associated with definite measurement
results, so that consciousness forks when a quantum measurement is
made.  Just as there is no consciousness whose experience is that of
the spacetime region occupied by you, me, Halley's comet, and the left
half of Georges Sand, so, after a measurement has correlated me with
the the z-spin of an initially x-polarized photon, there is no
consciousness whose experience is that of the full superposition (or,
once these branches of me are decohered, of the corresponding
mixture).  A more precise account of why would appear to await a
better scientific understanding of consciousness, though there are
probably some useful things to be said by philosophers, psychologists,
biologists, and decoherence theorists.  It is deeply bound up with the
problem of choosing a ``preferred basis'' in the relative state
interpretation (i.e., the question, ``relative to what?'', and also
with the problem of what tensor factorization of Hilbert space to
choose in relativizing states, which appears in this light as the
question of which subsystems of
the universe support consciousness).  One suspects the stability of
phenomena and their relations enforced by decoherence may underly the
ability to support consciousness, in contrast with a situation where
``Dr. Science'' is constantly reversing our seeming perceptions,
memories, etc...  Philosophers have devoted quite a bit of thought to
the conditions necessary for personal and mental identity to persist
through time (for things to be identified as manifestations
of ``the same person'' or ``the same mind'' at different times and places),
and this work is likely relevant here.  
Just why the branching should be described by a
stochastic process with probabilities equal to amplitudes squared is
somewhat mysterious; but I will not delve too far into it here; some
further thoughts on this matter, including an argument based on
Gleason's theorem (but which I now believe did not sufficiently 
motivate the assumption of noncontextuality) and some remarks on 
arguments based on tensor products and relative frequencies 
and an implicit assumption
that probabilities should be continuous in the state vector, can
be found in \cite{Barnum90a}.  \cite{Barnum90a} also develops my 
basic splitting-minds view of the relative state interpreation (which, 
I argue there, is a plausible interpretation of what Everett himself
had on the subject).  Several authors, notably Saunders \cite{Saunders98a}
and Wallace \cite{Wallace2002b,Wallace2003b} have given clear expositions of similar views.

Who is this ``Doctor Science'' I mentioned in the previous paragraph?
He is the modern version of Doctor Robert.  His powers are far greater
than Doctor Robert's; and he eschews Doctor ``take a drink from my
special cup'' Robert's crude methods, although he has refinements of
them available when needed.  To develop his methods, he had to perfect
another crude tool from the era of Doctor Robert: John Lilly's
isolation tanks.  He is able to isolate a human being completely from
interaction with his or her environment; preserving his or her quantum
state perfectly, and indeed, has universal, or at least extremely
fine, quantum control over the state of that human being.

Just such an image was probably part of David Deutsch's motivation for
investigating quantum computation.  He wondered what it would be like
if an ``information gathering and using device'' (IGUS, to use Murray
Gell-Mann's term) was correlated with something quantum (``measured
something on a quantum system'', ``gathered some information''), and
then, via Doctor Science-like quantum isolation and control, the two
branches of the wavefunction describing it experiencing different
measurement results were interfered in such a way as to reverse the
measurement (in an interview in 
\cite{Davies93a}).  Supposedly, the IGUS' report ``I experienced a definite
measurement result, but I can't tell you what it was'' would verify
the many-worlds (relative state) interpretation.  And just such an
image has guided many a defender of (or, as in my case, devil's
advocate for) the relative state interpretation, when asked such
questions as: ``what do you mean ``the different branches are all
real''.  Since we never see them, isn't it just a metaphysical
extravagance to suppose that they exist?''  A thought-experiment like
the one above involving Dr. Science
provides a kind of answer to this question.  We want to
say we have a superposition for the same reason as we sometimes 
say this about  atoms and photons: because, although in the case of a
human being the experiment is
impractically difficult, we could imagine having enough control over
the system and its environment, including the observer, to interfere
the different branches of the superposition, in such a way as to
either destroy the definiteness of the measurement result (measure
{\em on the system and observer} an observable complementary to the
one describing definite measurement results, say) or---here is where
control over the system's environment beyond just the observer is
crucial---disentangle the system from those degrees of freedom in
observer and environment that constitute a record of the measurement
result (possibly also making a joint system-observer-environment
measurement to provide some confirmation that this disentangling has
occured).  In the case of the relative state theory, though, the
system to be controlled might well be the entire universe, through
rapid decoherence.  This rapid decoherence, of course, is one of the
main things that makes Doctor Science's task so difficult.  And those
opposed to the RSI might say ``we can't get outside
the universe and interfere its branches, so the statement that it is
in a superposition is meaningless.''  Or if not meaningless, at any
rate superfluous, and therefore should not be included in a scientific
theory.  Now, the claim that this statement would be meaningless for
that reason 
seems to invoke a verificationist theory of meaning of a sort that, to
put it mildly, is now considered somewhat problematic by philosophers.
As to whether it is superfluous and therefore should be excluded from a
scientific theory, that is certainly far from clear.  The question, as
the more sober advocates of this superfluity claim
admit, is more one of judgement or taste.  Rather than showing that
the RSI is inconsistent, or impossible, it just displays features they
don't like, features they think are likely to prove impediments to
scientific progress.  I don't necessarily disagree.  I suppose my
situation is that I don't think the RSI is true; but I think pushing
it to the limit can be a very useful exercise in dealing with the
foundational problems posed by quantum mechanics: as useful as, and
indeed in a strange way part of the same project as, pushing to the
limit the view that ``quantum states are subjective.''  

One motivation for viewing superpositions in the universal state
vector as real is the ``slippery slope argument.''  Probably by
definition, doing interference experiments with the universal state
vector is impossible.  Indeed, when we don't have quantum mechanics
integrated with quantum gravity and cosmology, it is a bit of stretch
to suppose that such a thing as a state vector (or density operator)
of the universe makes sense (though one motivation for it was
precisely to start combining quantum mechanics and cosmology).  But on
the other hand, one can sometimes reverse model measurements in
microscopic systems, and if quantum error correction, noiseless
subsystems, active noise suppression, and fault tolerant computing and
control techniques which have been developed, mostly theoretically,
within QIS, should prove implementable, then one may have available
the tools to reverse measurement-like interactions involving far
larger and more complex systems.  Where does one draw the line?  The
subjectivist way is just to say, ``include whatever is appropriate for
this experimental situation, e.g. whatever you really have quantum
coherent control over.''  And, despite their willingness when pressed
to sometimes concede that they can't show the RSI is inconsistent,
they also sometimes claim that it is inconsistent for an observer to
view him or herself as described by quantum theory \citep{Fuchs2000a}.
Unfortunately, one doesn't see (at least I haven't seen) these vague
allegations of inconsistency made into a rigorous argument.  Even
doing so within some toy model would be valuable.  I think there is
likely a valuable insight here, but its value will not be made
manifest if no attempt is made to sharpen it.  Also, from the relative
states point of view, even if it is shown that it would be
inconsistent {\em for an observer herself} to have a complete
quantum-mechanical description of herself, the system she is
measuring, and the part of the universe that decoheres her ``in the
pointer basis,'' that of course does not show that such a description
is itself inconsistent.  The fact that it may be inconsistent, for
instance, to assume that we could have a completely detailed quantum
picture of the world (for one thing our brains are probably not big
enough, even collectively and supplemented by available computing
power), doesn't seem to me to say much about why {\em quantum}
mechanics is particularly ill-suited to play the scientific role, as a
theory in terms of which, say, ``God'' could formulate a description
of the universe and its dynamics, that people are supposed (on some
accounts) to have ascribed to classical physics.  Similar
``self-referential paradoxes'' \cite{Fuchs2000a} seem just as
threatening (and as potentially irrelevant) for a classical
description.

Another beef with the subjective point of view is somewhat akin
to Bell's beef with those who would take ``measurement'' as an
undefined primitive of quantum theory: there is too much vagueness, in
my view, about when it applies.  But I have hope that it can be made
less vague.  It could be that the difficulty, if it can be quantified
or otherwise better understood theoretically, of reversing
measurements made by complex systems such as organisms, is a deep
insight into why the RSI is not the correct ``overarching picture''
for physics.  Gravity could be involved: roughly because in current
theory there is not negative ``gravitational charge,'' it may be
theoretically impossible to shield things from gravitational
interactions, and these may be sufficient to decohere any
macroscopically different subspaces of a small subspace of the
universe, like those corresponding to different positions of a pointer
needle in a gauge.  Statistical physics seems potentially relevant,
too: quantum error-correction, and probably also active maintenance of
noiseless subsystems, has a thermodynamic cost \citep{Nielsen98a}, 
and this cost
 may become prohibitive for macroscopic systems.
But rather than just welcoming the ability to view quantum mechanics
as only appropriate to describing an observer's perspective on a
system, revelling in the subjectivity of it all, the way it perhaps
leaves room for mind, freewill, etc... as unanalyzed primitives, I
think it is still promising to try to get a grip on these matters
``from an outside point of view.''  I think it rather likely that the
relative state interpretation is wrong, and that quantum mechanics, as
the subjectivists insist, indeed is {\em not} appropriate for
describing the universe in terms of a universal state vector.  But the
attempt to understand its limitations in terms of a picture that
coordinates local perspectives, some of which are appropriately
described by quantum mechanics and some of which (because, perhaps, of
in-principle or maybe ``practical'' operational limitations imposed by
decoherence, the impossibility of gravitational shielding, or
whatever), need to be described differently, seems promising, and
related to serious physical questions involving statistical mechanics,
cosmology, and perhaps quantum gravity.
 
An analogy might be special relativity.  Here, an overarching picture
was achieved by taking seriously the fact that position and time
measurements are done via operations, from the perspective of
particular observers.  The heart of the theory is to coordinate those
perspectives into a global Minkowski space structure.  Certain aspects
of the local operational picture had to change in order to do this.
In particular, although one could still make the standard time and
position measurements according to standard procedures, the overall
picture implied limitations on the possible results of such
measurements: notably, velocity measurements will never come out
greater than the speed of light in a vacuum.   (Indeed, this limitation
was a motivation for the structure.) In a similar way, some
assumptions of standard quantum mechanics may turn out to be 
invalid, though hopefully in a way as gentle as the limitation
on velocities in special relativity, when the quantum-mechanical
perspectives are coordinated into an overall whole.  The alternative
would be the RSI, which I view as repugnant and also (as I suspect
do the subjectivists) as boring in something like the way that the Galilean 
coordination of local frames is boring compared to that of 
special relativity.  But I don't think that we should give up
on an attempt at such coordination, perhaps celebrating the
fact that quantum mechanics has shown us that it will be impossible
to achieve under the aegis of physics, just yet.  

\subsection{Beyond Hilbert space?}
An important point brought out by the attempt at a relative state
interpretation of quantum mechanics is the need to bring in, in
addition to Hilbert space, notions of preferred subsystems
(``experimenter'' and ``system'' perhaps also the ``rest of the
world'') or preferred orthogonal subspace decompositions (choice of
``pointer basis'' \citep{Zurek81a}).  It seems unlikely, as Benjamin
Schumacher likes to point out, that a Hilbert space, Hamiltonian, 
and initial state,  
will single out preferred subspace decompositions, hence the
relative state interpretation points out the need for links to aspects
of physics beyond Hilbert space.  \footnote{One wonders, though: suppose we
have $2^n$ energy levels such that there is a ``generic'' set of
positive reals $\epsilon_i$ such that the energies, indexed
by subsets $K$ of $\{1,...,N\}$,  have the form
$E_K = \sum_{i \in K \subseteq\{1,...n\}} \epsilon_i$; isn't the most compact
description of this situation just to say that we have $n$ two-state
systems, with energies $\epsilon_i$?  In the nongeneric case, say all
$\epsilon_i$ equal, though the spectrum including degeneracies would
be that of $n$ qubits, we would be unable to associate individual
energy states with ``which qubit is excited.''  (Put another way,
there would be many tensor product factorizations into qubits with
energies $O, \epsilon$ compatible with the Hilbert space and
spectrum.) With some degeneracy breaking, say of the form that
would come from small, unequal local perturbations
$\delta_i$ to the excited state energy levels of a bunch of qubits,
perhaps we could associate a ``best'' qubit decomposition with the
spectrum.  If this is
so, it suggests the nice idea that the existence of sufficient order
requires a tiny bit of disorder, to serve as a foil.  Perhaps in some
theories, this is the role played by initial conditions, versus laws.
The laws isolate the order; disorder is concentrated in the initial
conditions.}  
Schumacher has also criticized the RSI on the grounds that specifying
a Hamiltonian evolution on a Hilbert space can be made to look essentially
trivial (literally trivial, on a finite-dimensional space) by a 
time-dependent change of basis.  If one takes the view that ``the 
classical world'' is supposed to {\em emerge} from this structure
(Hilbert space, Hamiltonian, and initial state), then perhaps such 
transformations are legitimate.  On the other hand, this transformation
is not wholly trivial:  if one specifies a dynamics on a Hilbert space,
one is implicitly specifying a continuous 
canonical isomorphism at each time between
a continuum of isomorphic Hilbert spaces parametrized by $t$.  Then one
can in addition specify a Hamiltonian evolution (essentially, another 
such canonical continuous identification).   One can do this; {\em if}
it were to lead to something interesting as far as picking out a 
set of subspaces that are special with respect to this structure, that
would be interesting.  I have doubts this will work;  I also like Schumacher's
criticism that this specification of ``two connections on a fiber bundle
instead of just one''
seems mathematically unnatural.  But
I am not wholly convinced by Schumacher's 
criticism.  For one thing, I view the RSI less as a way of
getting the classical world of macroscopic objects and so forth to
emerge from Hilbert space, and more as a way of giving a realistic 
interpretation to Hilbert space structure in the presence of additional
structure such as preferred bases or subsystem decompositions
that represent other aspects of physics, without necessarily trying to derive 
these aspects from a Hamiltonian on a Hilbert space.
Schumacher views his argument against the RSI as also showing that one
needs these additional aspects of physics---{``nails in Hilbert space,''}
he calls them---to get a canonical identification of, say, bases from 
one time to the next (as, e.g., the spin-up/down basis). 
He interprets
this as showing the appropriateness of the Hilbert space description 
as applying to subsystems where the special structure lies in relations
to other systems (especially, for instance, macroscopic measuring devices,
such as a Stern-Gerlach apparatus which distinguishes an ``up/down axis''
in space), and the inappropriateness of the Hilbert space 
structure for the description of the whole universe.  
There are, of course, plenty of such non-Hilbert space 
aspects of physics, involving
symmetries, spacetime structure, and so forth.  Often, however, these
combine awkwardly with Hilbert space, as even quantum field theory
suggests (I refer to the need for renormalization in standard
Lagrangian quantum field theory approach), and as the difficulties
with attempts at quantum gravity make very clear.  The point is that
some of these
issues arise whether or not we take the relative state point of view
on foundations.  Yet, one hopes---at least I do---that this indicates
where to look for a resolution of the tensions between inside and
outside views--- that it is this difficulty in squaring quantum
mechanics with known physics, especially ``geometrical,'' ``outside''
aspects of physics, that suggests the distasteful aspects of the
quantum-mechanical outside view may vanish once such a squaring, with
whatever flexing is necessary from both sides, is accomplished.

\subsection{Beyond spacetime?}
I have said little about the Bohmian modification of quantum theory, 
but here is perhaps the place to remark that it is of interest as
an example of one of those nonlocal hidden variable theories
that Bell showed are the only non-conspiratorial way to realistically
model the statistics of quantum measurements.  Non-conspiratorial
refers to the prohibition, implicit in Bell's theorem and explicitly
mentioned by him elsewhere, on explaining the statistics of quantum
measurements by correlations between the hidden variables and what
we ``choose'' to measure.  In this regard, if one is thinking along
the lines that the lack of unification of quantum mechanics with general
relativity and gravitation may  lie behind interpretational problems, 
locality seems potentially a petty restriction:  when we are
contemplating quantizing the spacetime metric or otherwise 
unifying gravity and quantum mechanics, perhaps it is not
too farfetched to  imagine that spacetime and causality will
turn out to be emergent from a theory describing a structure at
a much deeper level....if this structure turns out to contain things
whose effects, at the emergent level of spacetime, can be interpreted
as those of ``nonlocal hidden variables,'' this should
hardly surprise.  Of course, I do not know what such a theory might
be like, and there is little reason to suppose 
its emergent description of induced ``spacetime hidden variables''
would resemble 
the specifics of the Bohmian theory, but it is useful to have at least
one concrete example (and it might be useful, for those who seriously 
contemplate a unification project going ``beyond spacetime'', to construct
other examples of, or investigate general features of, nonlocal hidden 
variable theories).

\subsection{Subjectivist and Objectvist views of the quantum state}

The other main attitude to quantum mechanics is to abandon the idea
that the quantum state describes reality, viewing it as a description
of our beliefs or knowledge about the outcome of experiments that
might be performed on a quantum system.  Here, too, there are
difficulties.  There is a drive to understand experiments that one
might perform as interactions between the system and some apparatus
coupled to us; many aspects of experimentation require adopting this
point of view, and then one could argue one is on a ``slippery slope''
leading towards viewing ourselves as quantum systems interacting with
the system being experimented on.

One might characterize the subjectivist attitude as an
``instrumentalist'' view of quantum theory: it is a procedure for
calculating the probabilities of measurement outcomes in experimental
or other observational situations, and the objects referred to in this
formal procedure should not necessarily be taken as ``elements of
reality.''  Note that this does not mean that quantum mechanics tells
us nothing about reality.  It tells us that reality is such that this
procedure works well, although not much about {\em why} it does.  But
it explicitly renounces the attempt to get a {\em physical} picture,
from ``outside'' the experimenter, of how and why this procedure
works, a picture which views the experimenter as part of a reality
which is at least approximately, or well enough for our purposes,
described by some mathematical structure, which is interpreted by
pointing out where in this structure we, the observers and
experimenters, show up, and why things end up looking as they do to
observers in our position.  It could be that the overall mathematical
structure is very different in nature from the descriptions observers
within the overall structure give of things from their perspective.
Or it could be, as in the relative state interpretation, that the two
kinds of structure are similar.  In my view, one should underestimate
neither the importance of pragmatic, predictive procedure, nor of
situating ourselves within a larger, at least partly comprehensible,
structure, as goals of science.  The subjectivist view might seem to
give up on the latter, while in my view the RSI shows that, at a high
cost, one can preserve both.

Despite the value I ascribe to situating ourselves within a larger
scientific picture of the world, I am not yet willing to believe in
the RSI.  But I will be clear about why: I just do not like the
picture it presents, of my consciousness constantly forking into
perhaps infinitely many branches, all of which have as much claim to
reality as the branch ``I'' who am writing this happen to be identified
with.  I don't like it because it seems to devalue the particularity
of the world as I experience it, instead injecting a disagreeable
element of ``everything goes'' into reality.  The negligible
probability  of bizarre quantum fluctuations rendering my world
apparently uncontrollable and weird (while still allowing me to remain
conscious) seems comforting, despite their theoretical possibility,
within a Bohrian interpretation of quantum mechanics; with the RSI,
all these bizarre possibilities are real, and the smallness of their
amplitudes in the world's state vector does not seem all that
comforting.  Perhaps to some my rejection of the RSI on these grounds
will seem like wishful denial of the hard truths of science.  This may
be reinforced by my attitude towards Schulman's proposed resolution of
the problems of quantum mechanics \citep{Schulman97a}.  As I
understand it, he proposes to retain essentially a one-Hilbert space,
state-vector evolving according to the Schrodinger equation,
no-collapse version of quantum mechanics, interpreted realistically if
you like, but to avoid macroscopic superpositions (Schr\"odinger cats) by
bringing in cosmology and statistical mechanics, and arguing that
symmetric consideration of final conditions along with the usual
initial conditions (that the universe was once much
denser and hotter than it is now) rules out macroscopic
superpositions.  There is a lot to do to make this persuasive, but
perhaps it can be done; it is certainly an ingenious and appealing
idea.  And if it does work, I am fairly happy to retain the rest of
the relative state metaphysics, now that I will not be committed to
the disturbing existence of forking D\"oppelgangers in subspaces of
Hilbert space decohered from me, but still real.  So that my discomfort
with the RSI can perhaps be summarized in 
a motto: ``Enough of this forking nonsense.''

\subsection{Mind and matter}

A more philosophical objection to the RSI is made by \cite{Bilodeau96a}.  In
some respects this is just the same thing, but in others, perhaps,
more substantive.  ``[The RSI's] purpose, however, is to maintain the
objectivity of a formally describable state of the universe as a
whole.  It does this by attributing the ambiguity that arises from the
``relative'' nature of the states of subsystems to the perceptions of
the observer.  In other words, the ``objective nature of physical
reality'' (that is, the formal independently-defined quality of the
systems described by physics, which are taken to be the ontological
basis of all reality) is maintained by shifting everything we think of
as objective physical fact (in the common intuitive sense of the
specific details of the world as we experience it) over to the
subjective side of the Cartesian split---that is, over to oblivion (as
a materialist undertands subjectivity).''

This contains the same distaste as my objections above: he doesn't
like to have the specific history of his experiences rendered merely
one among many perspectives.  But it also suggests a perhaps more
substantive, metaphysical objection.  The objection is that
fundamental to the RSI is the notion that consciousness and other
mental facts and things {\em supervene} on the physical.  The
``ontological basis of reality'' is taken to be a quantum system
described by a state vector.  A presumed philosophical and/or
scientific psychology, of the sort I have already mentioned, and whose
details perhaps do not yet exist, is trusted to explain why the
conscious perspectives on the world are as they are, and fork the way
they do.  Even before any consideration of physical theory, it seems
Bilodeau would object to the idea that consciousness supervenes on the
material world.  This is the point of his remark that the RSI consigns
this subjectivity to ``oblivion, for a materialist.''  However, I
think it is possible to maintain a version of the RSI which is not
materialist in this sense: the notion of supervenience of mind on
matter, in contradistinction to ideas of identity between them (or
bogusness of one or the other), is around precisely to allow such moves.
Bilodeau seems to view supervenience as almost as bad as oblivion.
But it seems to me that the RSI is also consistent with a ``dual
aspect'' theory of the mental: instead of saying the mental supervenes
on the physical, arising like the froth on a wave whenever certain
physical conditions are met, we might more congenially say it is
another, subjective, aspect of things, things which can also have
physical or objective aspects.  The subjective aspect is
what it is like to be those things
(\cite[][Chs{.} 12, 13]{Nagel79a}, \cite[][esp{.}  Chs. II,
III]{Nagel86a}).  In some cases, the physical system becomes complex
enough, or whatever, that ``what it is like to be that system''
becomes substantial enough to be called a mental phenomenon.
This does seem preferable to supervenience, which makes mental
phenomena seem somehow superfluous, just along for the ride but not
part of the essence of things.  But the difference between
supervenience and ``subjective aspect'' may be less than it seems.  To
make it substantive may require a move which Bilodeau wants us to
make: to take the view that we will not understand the subjective
aspect of things in terms of physical theory, so that the mental
aspect of things takes on a life of its own, rather than just being an
aspect of something adequately understood in terms of the physical.
However, I would not like to make too many assumptions in advance
about the terms in which we will (if we will) ultimately be able to understand
things about consciousness and the mental.  Almost certainly it will
be in terms of concepts, perhaps scientific concepts, at a much higher
level than those of physics: this is familiar from sciences like
biology and chemistry.  I just don't want to assume that understanding
enough about mind (which may not be very much; it may even be
compatible with the impossibility of a physicalist understanding of
many important aspects of the mental) to have a stylized psychology
capable of explaining the perspectives, and the forking of
perspectives, required by the RSI, requires a thoroughgoing
ontological committment to the ``primacy of the physical'' in some
sense.  Indeed, even if physical terms (or higher-level terms no more
``unphysical'' than those of chemistry or biology) suffice for a
reasonable {\em understanding} of mental phenomena, or many aspects of
them, I'm not sure this would rob them of {\em ontological}
independence, in a dual-aspect theory.

Of course, Bilodeau doesn't very explicitly make the argument I have just
made, that the RSI's possible 
need for a materialist psychology might be problematic
enough to make us consider alternatives.  In any case, I am arguing
that an 
RSI-like gambit can be run even by someone who believes in the reality and
substantial independence of the mental aspects of things;  the relation 
of physical systems to consciousness will be subtler perhaps in such a theory,
but it is hard to imagine there won't be some such relation capable of 
sustaining an RSI-like story.  The bottom line, I think, is really the 
matter of preference, even ego, I spoke of above.  
We just don't like the fact that the RSI dethrones {\em our} particular
perspective from its privileged position, and establishes alongside it
a possibly infinite number of variants of it, 
some of which (although with negligible amplitude)
involve bizarre and fantastically improbable events, 
along with many other 
perspectives which are not even those of anything we think of as existing
in our world.  But again, perhaps that's just the way things are.
Other branches of physics have suggested, with various degrees of 
plausibility, that there is stuff outside our forward and backward
light-cones, maybe even completely causally disconnected from our forward
light-cones, maybe even trapped in a bubble in spacetime called a 
``baby universe;''  if so, there may well be stuff there that is 
organized in such a way as to have consciousness, and that's just how
science tells us it could be.  But perhaps what is objectionable about
the other Hilbert subspaces 
of the RSI and not about these inaccessible parts of spacetime, is 
the wildly implausible nature of the things that happen in them.

What appeals to Bilodeau, I think, about the Bohrian interpretation is
that it suggests that science is {\em showing} us that the
epiphenomenal (supervenient) ``materialist'' theory of mind I
mentioned in connection with the RSI, is untenable.  One could, of
course, adopt a wholly instrumentalist view of quantum theory, in
which case it would presumably not be telling us anything at all about
mind.  But Bilodeau wants to argue that the very fact that this procedure
works tells us something about reality.  The fact that attempts to
describe it, and our place within it, from an overall point of view in
a physical theory, supposedly fail just when we try to place
ourselves, and our experiences of definite quantum measurement
results, within it, supposedly shows that our mental aspects such as
capacity to have experiences cannot be incorporated into an
empirically accurate physical picture of the world, and so {\em must}
be given a certain degree of autonomy and fundamental role in an
adequate picture of reality.  Such a picture is 
fairly eloquently, if necessarily
vaguely, described on p. 400 of his paper.  ``There is Being.  Being
is aware.  Being acts.  The action of Being (from our perspective as
participants) represents itself (in part) as the physical universe in
historical space and time.''  The main point is that
``...consciousness must be closely related to existence itself, that
it is vastly nearer to the ``basic level of reality'' than anything
signified by physical concepts.''  I find these notions enormously
attractive both on philosophical grounds and at a pre-philosophical
level.  Yet I would not want to prejudge the potential of scientific
consideration of mental phenomena because of this attraction.  And I
also am not as convinced as he is that quantum theory has decisively
cast the die in favor of this kind of view, which may have to survive
a bit longer on its own intrinsic merits.  Indeed, I am not convinced,
as suggested above, that the RSI does not provide a workable, even if
unpalatable, way to do just what Bilodeau claims quantum physics
forces us to renounce: view ourselves as part of a larger reality,
about which reality it is not clear we need make, on physical grounds,
a judgement about the degree to which it is ``fundamentally
physical,'' or can be understood in physical terms.  In other words, I
think the RSI can be made to work to some extent {\em independently}
of how tightly consciousness is linked to or subordinated to the
physical, as long as the linkage is sufficient to get the various
needed subjective perspectives working right.  It thus seems possible
to me that the unpalatability of the RSI's picture of what minds exist
is just one more sign of the unfinished scientific business of
integrating quantum mechanics with the rest of physics.

We would like to take a view in which stuff happens to us, and this 
``historical'' account of what happens, the incontrovertible stuff of
everyday reality, is made sense of by finding regularities in it, 
regularities which turn out not to be tight enough to be describable
by deterministic causal laws, but still tight enough to be describable by
stochastic laws.  The resistance to the RSI expresses a reluctance to 
dethrone this ``historical'' account of what happened to us, to a par
with other experiences going on in other branches of Hilbert space.
Yet, this is the sort of thing that science often forces us to do:
acknowledge the existence of stuff---real, physical stuff, but also 
stuff which it is only reasonable to expect in some cases 
has as much ``subjective,'' ``historical'' existence as our experiences:
other planets, other galaxies, perhaps wildly remote regions of the 
universe with which we will never be able to communicate in 
our lifetimes.  

Further investigations from the point of view of the RSI seem to me just
as justified, even as likely to be fruitful, as those from the 
``subjective''  point of view.  Indeed, I have argued that to a certain 
extent they may closely parallel each other, inasmuch as the RSI also 
takes the probabilistic interpretation of measurement outcomes and
statevector collapse as ``subjective,'' appropriate for understanding
how things are likely to look from a certain perspective.  

\iffalse
[Bilodeau:  on seeming resolution of paradox.  But then:  slippery
slope (of what is subjective; of the ``cut'').  
Later:  geometry vs. dynamics.  B inadequately deals w/ 
relativity.]
\section{}
\fi

\section{Laws of thought and overarching structures}
\label{sec: laws of thought}

\iffalse
[Stuff on QM as a law of thought?]
\fi

Partly because in 
my view it has not yet resulted in significant substantive progress
in understanding foundational problems of quantum mechanics,
I think the most important impact of the emergence of QIS
on the foundations of quantum mechanics is
to provide a formal framework for analyzing and making quantitative
those problematic or peculiar aspects of quantum mechanics,
in the guise of the theory of information (broadly understood).
The most significant potential
impact of this information formulation 
lies in the fact that information is being recognized (and this
simultaneously with, and to a fair extent independently from, the
emergence of QIS) as critical to the understanding of many aspects
of physics.  This is most notable in statistical mechanics.  
The foundations of statistical mechanics 
have also been somewhat controversial;  perhaps less
so than those of quantum mechanics, but for some of the same
reasons:  it is difficult to sort out the subjective
and objective aspects of the foundations of the theory.  In my
view, the foundations of statistical mechanics are probably
best understood in at least partly subjective terms:  a system
has high entropy for an observer if that observer lacks information
about it;  information is power, in the sense that an observer
having more information about a system, and hence a lower entropy
for it, is likely (other matters equal) to be able to extract
more useful work from it.  This view emerges from the analyses
by Landauer, Bennett, Zurek, Caves and probably others, of the
Maxwell Demon paradox, and it shares with the subjective interpretation
of quantum 
mechanics the idea that the theory (be it statistical mechanics or QM) is
suited for describing one part of the world as seen by another, 
not necessarily giving a model for the world as it might be 
seen ``from outside,'' independent of 
any viewpoint within it.

Yet the issue of subjectivity seems somehow more pressing for 
quantum mechanics, because one has gotten used to the idea
that statistical mechanics is not a ``fundamental theory,''
but an emergent aspect of a world governed by fundamental 
laws, indeed to the idea that certain aspects of statistical
mechanics will likely be emergent from fundamental laws of
many different forms, so that although in one sense less
fundamental, they may actually be more universal than 
particular theories of the fundamental laws.  Perhaps the
information-theoretic view of quantum mechanics could
be interpreted as saying that actually, QM is similar
to statistical mechanics in this regard:  QM should be
viewed as a theoretical framework appropriate for the
description of one part of the world by another, and not
for the description of the world ``from outside.''  Yet
this is compatible with the idea that quantum mechanics
will be seen to emerge from a more fundamental theory...
which smacks of hidden variables.  It would be
interesting if QM had some ``universal'' properties similar
to those which statistical mechanics probably has...i.e.,
that quantum mechanics ``must'' emerge as a theory describing
the perspective of one part of the world on another, whenever
the ``underlying'' theory has certain properties.  This is
reminiscent of, though far from the same as, the idea that
``quantum mechanics is a law of thought.''  The two ideas may
be brought into closer relation
if we take this last statement not as obviously analytically true,
(in the sense in which some might wish to say that it is 
analytically true that ``Bayesian 
probability theory and/or decision theory is a law of rational thought''), but
in a somewhat Kantian sense, that the aspects of the world which
give rise to quantum mechanics as an emergent description of
perspectives within the world, are necessary for the existence of
rational beings having perspectives (or in a weaker version,
perspectives of a certain sort). 
Though I have no particular model of 
how this could be true, it
seems quite plausible that a certain degree of regularity 
in the world is necessary for the emergence of rational,
indeed probably even sentient, beings,
and conceivable that a certain combination of regularity and 
irregularity, manipulability and uncontrollability of parts
of the world by others, which is just right for a 
quantum-mechanical description, might be necessary for this emergence
of rational and/or sentient beings.

Combined with an argument for the unknowability, even perhaps
the pointlessness of attempting approximation via science, 
of the ``underlying''
description, this might begin to be a possible interpretation of
the information-theoretic take on the foundations of quantum 
mechanics.  More radically, one might wish to combine it with an
argument for the ``nonexistence'' of this underlying description.

\section{The combination of perspectives}
\label{sec: combination of perspectives}
The viewpoint I am advocating, then, is that we should continue to
maintain both the inside and outside view of quantum systems, and in
interpretational matters to pursue a better understanding both of the
possibility of viewing quantum theory as about the dynamics of
information-like, perhaps subjective, states, and of the possibility
of viewing it as about the sorts of entanglement and correlation
relations that can arise between systems.  A prime example of a
worthwhile problem along the former lines is that being pursued by
Caves, Schack, and Fuchs, of just how far one can push the analogy
between quantum states and subjective Bayesian probability
distributions, in particular how quantum measurements and ``collapse
of the wavefunction'' can be viewed as similar to the updating of a
Bayesian probability distributions upon gaining new information.  A
prime example of a worthwhile problem along the latter lines is
understanding how the probabilities for collapse can be understood
within the RSI, again as something like (or perhaps as exactly) a
Bayesian updating process of ``gaining more information about which
branch of the wavefunction we are in.'' As I just phrased it, this
sounds {\em exactly} like Bayesian updating, but the key potential
difference is that we are not gaining information about a pre-existing
fact, but rather about a ``perspectival'' fact (such as ``I am in the
spin-up branch of the wavefunction'') that is not even true or false
before the measurement procedure creates the branches.

But equally, or perhaps more, important problems can be posed without
committing to either point of view, and it is this sort of problems,
and the role of quantum information and computation in the project
they suggest, that I want to emphasize here.  I suggest that
the ``operational'' or ``empirical quantum logic'' point of view is
useful whether one wants to consider the empirical theory as for some
reason all we can hope for, or as a description of how perspectives
look within an overarching theory such as the RSI.  Indeed, the 
similarity between the problems posed above is an example of how
the operational approach is relevant to both:  investigate quantum
mechanics' properties as a theory of perspectives of subsystems
on other systems, without prejudging whether or not the perspectives
will turn out to be coordinatable into an overarching picture---indeed,
while trying to ferret out how this might happen or be shown to be
inconsistent, and how this possibility or impossibility may be
reflected in the operational, perspective-bound structures.  

This general operational approach {\em may} turn out to be a limiting,
or limited, way of viewing physical theories.  It seems rather suited,
however, to quantum mechanics, particularly Copenhagen-style or
``subjective-state'' quantum mechanics, and also adequate to treat
classical mechanics in its information-processing aspect.  The
limitations arise because this sort of theory takes ``operations,''
such as making a measurement on a system, as basic terms in
formulating a physical theory.  It views a physical theory as a
description of the behavior of a ``system,'' a part of the world
viewed as susceptible to the performance of operations on it by,
presumably, another ``part of the world,'' the observer or
experimenter (\cite{Foulis98a}, p. 2).  However, it need not
necessarily be associated with a renunciation of the attempt to view
the entire world as one structure, and physical theory as describing
that structure.  It might be possible to also view the entire world 
in terms of an empirical theory of the same kind, from the point of
view of a fictitious observer.  (I was somewhat surprised to hear
David Finkelstein defend such a point of view a few years ago.)
Classical theory may be such a case, while many believe that quantum
mechanics is not.  However, Everett's ``relative-state''
interpretation (the inspiration for what is sometimes called the
``many-worlds interpretation'') attempts such an ``objective''
interpretation of quantum mechanics.  Indeed, Rovelli and Smolin
\citep{LSmolin95a} have proposed making the impossibility of such an
external view a desideratum for an acceptable physical theory.

But (aside from quantum gravity aspects) the main interest the
Rovelli-Smolin
approach holds for me is, 
that it suggests a framework in which 
quantum mechanics is good for describing things from the point 
of view of subsystems, but not appropriate for the entire universe, 
but in which {\em nevertheless} there exists a mathematical 
structure---something like a topological quantum 
field theory (TQFT), spin network, or spin foam---in 
which these local subsytem points of view are coordinated
into an overall mathematical structure which, while its terms may be
radically different from those we are used to, may still be viewed
as in some sense ``objective.''  However, it is far from clear yet
that this can be done while avoiding the more grotesque aspects
(proliferating macroscopic superpositions viewed as objectively
existing) and remaining conceptual issues (how to identify a preferred
tensor factorization, and/or preferred bases, in which to identify
``relative states'') of the Everett interpretation.  

In TQFT's or spin networks and generalizations, 
the description appropriate to ``perspectives'' is still
Hilbert spaces, but only in special cases do these combine as tensor
products.  If we view a manifold as divided into ``system'' and
``observer'' via a cobordism, then as the ``observer'' gets small
enough, while the ``system'' gets larger, we start getting, not the
increase in Hilbert-space size to describe the system that we might
expect as the system gets larger, but a decrease in Hilbert-space size
whose heuristic interpretation might be that the observer has gotten
so small that it no longer has the possibility of measuring all the
operators needed to describe the ``large'' Hilbert space one might
have expected.  The Hilbert space does not describe the ``large'' rest
of the world; it describes the relation between a small observer and
the larger rest of the world.  It would be nice if this could somehow
be made to imply that the observer cannot get into a 
Schr\"odinger cat-like macroscopic superposition, but it is not clear
this can be done.  Indeed, when I sketched this viewpoint to Carl
Caves, he asked how it differs from the Schmidt decomposition.  View
the entire universe as in a pure state (say $2^n$ dimensional).  As we
include more and more of the universe in one or the other of the
systems, the local supports of the state become smaller and smaler:
this can be interpreted precisely as I proposed interpreting the
reduction in Hilbert space size in TQFT's above: as a reduction in the
local Hilbert space size needed to describe local operations, given
the pure state that describes the system.  However, here all the
``apparatus'' etc... needed to perform local measurements shoud be
included on the observer side, so that in some sense, the
``perspective'' should be one in which a particular measurement is
perhaps committed to, or perhaps better, in which we use a
``decoherent histories'' kind of approach to ask which histories are
consistent with this pure state of the universe, this split into
observer and observed, and whatever dynamics one introduces.  And
indeed, Rovelli and Smolin found themselves moving toward the
decoherent histories point of view to interpret their theory, though 
Smolin at least felt that the Dowker/Kent objections to consistent
histories were a serious problem.  A more
``operational'' point of view, emphasizing a certain ``freedom'' to
set up different local experiments, might need to divide the manifold
into more parts, and specify the state less precisely than assuming a
single pure state of the whole.  Indeed, unless something like this is
done one anticipates difficulties with the usual assumptions of the
theory of quantum operations.  If the entire system is in a pure
state, or even certain types of mixed state, it may not make sense to
assume that our local apparatus is in a state independent from that of
the system.  Understanding when this sort of assumption can be
reasonable is a potential contribution of a theory of this sort.
However, this also makes the interpretation of the local Hilbert space
structure in the case where observer of system approaches being the
whole universe, problematic.  I do not view such potential
difficulties, however, as reason to abandon the project of attaining a
view of the world as an overarching structure which coordinates many
perspectives.  Rather, it suggests reasons, more physical than
metaphysical, why the attempt to force that overarching structure into
a mold derived from standard quantum mechanics may fail.  But we need
to try to patch up these failures, modifying how systems combine, and
perhaps even introducing new categories of empirical theory, in order
to make progress toward such an overarching structure.  We may find,
as some in the 
the ``quantum states are subjective'' crew might anticipate, that
such progress is not possible, or not possible from the point of view
of physics alone.  This would be fascinating and significant, but I do
not think we are yet near the point where one should give up on the
project and accept its impossibility.  But even if quantum states are
subjective, this doesn't mean they can't be combined, as states from
perspectives, in a coherent way.  And whether or not we accept
the the task of attempting to make progress toward such an overarching
theory, we need to understand quantum mechanics, and classical
mechanics, as examples of empirical-theory types: how the distinction
between the two can be characterized, perhaps axiomatically; whether
we can obtain such characterizations in which the axioms have an
intuitive meaning, so that we understand the difference between the
two in terms of that meaning; whether both share certain features,
perhaps describable in terms of their power to process information,
that are not necessary features of empirical operational theories in
general; and so forth.  Here we might see how the quantum description
of certain perspectives could arise as a limiting case of some more
general type of perspective, which necessarily also arises in an
overarching structure that includes quantum-mechanical perspectives in
a physically reasonable way.  Or we might see how a non-tensor product
law of combination of subsystems---quantum or not---could be relevant
in some situations.  This is just the sort of thing that operational
quantum logic aspires to investigate.  The role of quantum information
and computation in this project could be of great importance: to
suggest axioms with clear meaning in terms of information processing,
and to elucidate their connection with other areas of physics such as
statistical mechanics or cosmology.  More mundanely but no less
importantly, operational quantum logic
could have a role to play in quantum information and computation
proper: particular suggested implementations of quantum information
processing, or other particular operational situations, which impose
limitations on the measurements and other operations that can be
performed, may correspond to structures of a kind already investigated
in some branch or branches of operational quantum logic, 
which may supply mathematical tools
and results with which to analyze them.

\section{Frameworks for empirical operational theories}
\label{sec: operational theories}
\subsection{Introduction:  operational quantum logic and convex sets frameworks}
In this section I will introduce frameworks, 
that I find particularly useful for thinking
about empirical operational theories.  The general area of
mathematical descriptions of operational theories (``mathematical
metascience,'' as David \cite{Foulis98a} has called his (and collaborators')
version of it in an
excellent introduction and overview) seems to me to exhibit some
particular mathematical behaviors that it can help to take note of
before plunging into the literature.
One is that many constructions can be described in alternate ways,
notably involving the relationship between order-theoretic and
algebraic descriptions of the same thing, so that some of the
most interesting categories have been independently defined in different
ways.  A notable example is ``effect algebras,'' introduced under
this name by Foulis and Bennett \citep{Foulis94a}, but, it turns out,
introduced as ``weak orthoalgebras'' in
\cite{Giuntini89a}, and independently, from a more order-theoretic
point of view, as ``difference posets'' (D-posets, for short) by
\cite{Kopka94a}.  Also there are frequently nice functorial relations
between different categories of empirical theory types, sometimes inducing
equivalences under mild conditions, again providing different
perspectives on the same structures.  A notable example is the
representability of ``convex effect algebras'' on ordered vector spaces
\citep{Gudder98b,Gudder99a}, and related results.  The precise
statement is that any convex effect algebra admits a representation as
an initial interval of an ordered linear [i.e. vector]
space.  Long before this representation theorem was proved,
effect-algebra-like structures had been noticed in the convex
state-space approach to operational quantum mechanics and more general
operational theories \citep{Davies70a, Ludwig81a, Ludwig83a}, and
indeed, the term ``effect'' was introduced (so I have been told) by
Ludwig, who worked within a version of the convex states approach.
That some such relation existed has been suspected for quite awhile
(see, for example \cite{Cook81a} on the relation between the convex
approach and a predecessor of the effect algebra approach); also,
similar linearization theorems exist for other types of quantum
structures (R\"uttiman, cited in \cite{Gudder98b}).

Another notable general feature of the mathematics of this area is 
a certain ``smoothness'':  it is usually possible to make small 
changes to the axioms describing a category of operational theories, 
and the theory of the category changes incrementally rather than 
radically as a result of such changes.  (No doubt someone will point to
some welcome counterexamples;  but there is a lot of
such ``smoothness,'' overall.)  This results in a profusion
of results, often closely related but slightly different and sometimes
stated in 
very different terminology.

The result of these general features is a body of mathematics which is
rich and fascinating, but which also lends itself to duplication of
terminology and results, and a certain amount of resulting confusion.
Often a researcher will find it easier to develop results closely
related to another's, within his own framework, rather than make the
effort necessary to learn yet another set of notations, results, and
quirks.  Yet one expects---and this is occurring---that in the end, it
will be useful to develop a broad view encompassing many different
approaches to quantum and other empirical ``logics'' or
``structures.''

\subsection{Frameworks for representations of empirical probabilities}

My preferred approach to empirical or operational theories is to start from
the compendia of probabilities that our phenomenological theory says
are possible for the different possible results of different possible
operations on a system (indeed, I will often call these compendia
phenomenological theories), and attempt to construct various more abstract
structures for representing aspects of empirical theories---effect
algebras, classical probability event-spaces, $C^*$-algebraic
representations, spaces of density operators on Hilbert spaces,
orthomodular lattices, or what have you---describing types of
operational theories, from these.  With most such types of abstract
structures, the possibility of constructing them in a specified way
from compendia of phenomenological probabilities will impose
restrictions on these compendia of probabilities, and the nature of
these restrictions constitutes the empirical significance of the
statement that our empirical theory is of this type (has this abstract
structure).  This approach promises to systematize our understanding
of a wide range of empirical structures and their relationships, both
mathematically and in their empirical significance.  Of course, a
large part of the existing body of work on empirical theories may be
viewed as part of this project, since the relationship to the
probabilities of experimental outcomes has always been a critical part
of understanding these structures as empirical theories, and even as
abstract mathematical structures, the space of ``states'' on such
structures is often a crucial aid to understanding their mathematical
structure.  This is of a piece with the situation in many categories
of mathematical objects.  $[0,1] \subseteq \R$ 
is a particularly simple example of
many categories of ``empirical structure,'' and a state is a morphism
onto it; understanding the structure of some more complex object in
the category in terms of the set of all morphisms onto this simple
object is similar to, say, understanding the structure of a finite
group in terms of its characters (morphisms to a particularly simple 
Abelian group, $S^1$).

In this project, I like to make use of an idea which has come in for a
fair amount of criticism, but has been with us from early in the game.
It certainly appears in Mackey's classic work on characterizing
quantum mechanics axiomatically \citep{Mackey63a}, but \cite{Cooke81a}
even ascribe it to Bohr (an ascription which I have not independently
checked).  It is also fundamental to Ludwig's convex approach to
foundations, and I expect the use I make of it may turn out to be
closely related to the role it plays in Ludwig's work.  One also finds
it on p. 14 of \cite{Mielnik69a}, and doubtless in many other places.
This is the notion of ``probabilistic equivalence'': two outcomes, of
different operational procedures, are viewed as equivalent,
``identified'' in some abstract sense (best interpreted, perhaps, as
``exhibiting the same effect of the system on the observing
apparatus,'') if they have the same probability ``no matter how the
system is prepared,'' i.e., in all admissible states of the
phenomenological theory.  The interpretation as ``exhibiting the same
effect of system on apparatus,'' is probably traceable to Ludwig, and
is perhaps  the motivation for his introduction of the term ``effect''
for the abstract entities that, in quantum mechanics, correspond to
the equivalence classes of outcomes under this relation.  
It also helps forestall the
main objection to this notion, which is that two outcomes equivalent
in this sense may lead to different conditional probabilities
(conditional on the outcome in each case) for the results of further
measurements.  One may say that they are equivalent only as concerns
the effect of the system on the apparatus and observer, not vice
versa.  This discussion implicitly supposes a framework in which
operations may be performed one after the other, so that outcomes of
such a sequence of $N$ measurements are strings of outcomes $a_1 a_2
... a_N$ of individual measurements.  Then a stricter notion of
probabilistic equivalence may be introduced, according to which two
outcomes $x$ and $y$ are equivalent if for every outcome $a,b$ (which
may themselves be strings), the probability of $axb$ is the same as
that of $ayb$, in every state of the phenomenological theory.  I
believe that in 
the quantum case, there is a bijection between the
equivalence classes of this relation and the trace-nonincreasing
completely positive maps (quantum operations).

The main point, of course, will not be to exhibit another formalism
with which to describe quantum mechanics, but to exhibit quantum
mechanics as a
special case of a more general type of empirical structure.  Then,
within the theory of such more general structures, one may investigate
the effect of imposing additional axioms, the information-processing
power of empirical theories more general than quantum mechanics, the
way in which such structures may combine as subsystems of a larger
system, and so forth.  Before dealing with the probabilistic
equivalence classes (often also caled ``operations'') on 
phenomenological theories in which we take account not just of single
measurements but sequences of measurements made in succession I will
deal with the case where the probabilistic equivalence relation
ignores sequential considerations (calling the resulting equivalence
classes ``effects'').  The resulting structures are likely still relevant
to general empirical theories (just as POVMs are still useful in
quantum mechanics even though in some cases one needs the additional
detail about conditional dynamics provided by ``instruments,''
collections of quantum operations that sum to a trace-preserving one,
terminology that may have been introduced in \citep{Davies70a}, and is
used also in e.g. \cite{Holevo98a}.)

\subsection{Effect algebras and related structures}

Before considering in detail the derivation of the structure of the
set of probabilistic equivalence classes (``effects'') of an
operational theory, I will introduce some of the abstract structures
we will end up with: effect algebras and ``weak effect algebras,''
motivating them (in the case of effect algebras) with classical and
quantum examples.

\begin{definition}
An {\rm effect algebra} 
is an object $\langle \ce, 1, \oplus \rangle$, where
$\ce$ is a set of elements called effects, $1 \in \ce$, 
and $\oplus$ is a partial binary operation on $\ce$ which 
is (EA1) strongly commutative and (EA2) strongly associative.  
The qualifier ``strongly,'' which is not redundant only because
$\oplus$ is partial, indicates that if the sums on
one side of the equations for commutativity and associativity
exist, so do those on the other side, and they are equal.
Explicitly, these equations are:
\beqa
\label{commutativity} &{} & \text{Commutativity } a \oplus b = b \oplus a\;\\
\label{associativity} & {} & \text{Associativity } a \oplus (b \oplus c) = (a \oplus b) \oplus c\;.
\eeqa

In addition,
\\
(EA3) $\forall a \in \ce, \exists ! b \in \ce ~~(a \oplus b = u)$.
(The exclamation point indicates uniqueness.
We give this unique $b$ the name $a'$; it is also called the
{\em orthosupplement} of $a`$.)
\\
(EA4) $a \oplus 1$ is defined only for $a = 1'.$  (We will often call
$1'$ by the name ``0''.)
\end{definition}

If we require that the equalities 
\ref{commutativity} and \ref{associativity} 
hold only when both sides are
defined, allowing the possibility that 
one is defined while the other is 
not, we call these ``weak commutativity'' and ``weak associativity.''

In the effect algebra of quantum mechanics (on a finite-dimensional 
Hilbert space, say), $\ce$ is the 
unit interval of operators $e$ such that 
$0 \le e \le I$ on the Hilbert space,
$\oplus$ is 
ordinary addition of operators restricted to this interval
(thus $e \oplus f$ is 
undefined when $e + f > I$), $1$ is the
identity operator $I$, and $e' = I - e$.    $0$ turns out to be
the zero operator.
Classical examples
also exist.  If we consider the set $\cf$ of 
``fuzzy sets'' on a finite
set $\Lambda = \{\lambda_1,...,\lambda_d\}$ (which are 
functions from $\Lambda$ to $[0,1]$), and define $\oplus$ 
as ordinary pointwise addition of functions (i.e. defining 
$f+g$ by $(f+g)(x) = f(x) + g(x)$ except that $f \oplus g$ is 
undefined when $f + g$'s range is not contained in $[0,1]$), 
and $1$ is defined to be the constant function whose value is $1$, 
then $\langle \cf, 1, \oplus\rangle$ is an effect algebra.
This algebra is obviously
isomorphic to the restriction of the quantum effect algebra on a
$d$-dimensional Hilbert space to effects which are all diagonalizable
in a particular basis.  These ``fuzzy sets'' may be interpreted as the
outcomes of ``fuzzy classical measurements'' in a situation where there are
$d$ underlying potential atomic ``sharp'' measurement results or
``finegrained outcomes,'' but our apparatus may have arbitrarily 
many possible meter readings, connected to these ``atomic outcomes''
by a noisy channel (stochastic matrix of transition probabilities, 
which are in fact the $d$ values taken 
by the function (effect) representing a (not necessarily atomic)
``outcome''.).

If we add the requirement
\beqa
(OA5) x \oplus x \text{ exists } \Rightarrow x = 0\;
\eeqa
then we have an {\em orthoalgebra};  the projection operators 
on a finite quantum system are an example.

\iffalse
Although both of these can be represented by in terms of $C^*$ algebras
(a full matrix algebra in the first case, a commutative one in the
second), the effect algebra structure is much weaker than the
$C^*$ algebra one, and many other examples of effect algebras exist
which cannot be represented in this way. 
\fi

Later, we will also be concerned with something I will call a ``weak 
effect algebra.''  This satisfies all the axioms above except that
strong associativity (EA2) is replaced by weak associativity (WEA2).
Obviously,
every effect algebra is a weak effect algebra, but not vice versa.
A strengthening of the effect algebra notion, 
called an {\em orthoalgebra}, is also of interest:
it adds the axiom that $x \oplus x$ exists only for $x=0$.  
The projectors on a quantum-mechanical system, with the same
definitions of $1, \oplus$ as apply to more general POVM elements,
are an example (as well as being a sub-effect algebra of the 
full quantum mechanical effect algebra of POVM elements).

Other weakenings of the notion of effect algebra have been considered.
Notably, Wilce considered a fairly general notion of ``partial abelian
semigroup,''  (PAS) which is like an effect algebra except that
we retain only (EA1) and (EA2); various combinations of additional
requirements then give a remarkably wide variety of algebraic structures
that have been considered in operational quantum logic, including effect
algebras, test spaces, E-test spaces, and other things.
In particular, an effect algebra is a positive, unital, cancellative,
PAS.  
Since we weakened (EA2) to obtain the notion of WEA, 
WEA's are not a subclass of PASes.  However, later, when we
introduce the notion of ``operation algebra'' we will want to 
generalize the notion of WEA analogously to the
generalization of EA's to PASes, by also weakening (EA3) and
modifying (EA4) to reflect this.

A {\em state} $\omega$  on a weak effect algebra $\langle \ce, \oplus, 
 1 \rangle$
is a function from $\ce$ to $[0,1]$ satisfying:
\beqa
\omega(a \oplus b) = \omega(a) + \omega(b)\;,
\omega(1) = 1\;.
\eeqa
A {\em finite resolution of unity} in a weak effect algebra (to be interpreted
as the abstract analogue of a measurement) is a set $R$ 
such that $\oplus_{a \in R}  a = 1$.   
So for a resolution of unity $R$, 
$\sum_{a \in R} \omega(a) = 1$: the probabilities of measurement results
add to one.   
A {\em morphism} from one WEA $\ce$ to another $\cf$
is a function $\phi: \ce \rightarrow \cf$ such 
that $\phi(a \oplus b) = \phi(a) \oplus \phi(b)$;  it is called 
{\em faithful} if in addition, $\phi(1_\ce) = 1_\cf$, where $1_\ce$ and
$1_\cf$ are the units of $\ce$ and $\cf$.   $[0,1]$, with $\oplus$ addition
restricted to the interval, is an effect algebra, so a state on $\ce$ 
is a faithful morphism from $\ce$ to this effect algebra.

I will attempt to avoid issues involving effect algebras and WEA's
where $\ce$ is infinite and infinite resolutions of unity are defined,
even though even finite dimensional quantum mechanics is properly done that
way.  To this end I will assume that EA's and WEA's are {\em locally finite}:
resolutions of unity in them have finite cardinality.
For example, for finite $d$-dimensional quantum mechanics, most
things should work the same if we restrict ourselves to work with
resolutions of unity into $d^2$ elements.  (We certainly need $d^2$
to provide a basis for the Hermitian operators on the Hilbert space.
Local finiteness, of course, is much weaker than the existence of such
a uniform bound.)
Those who are interested in structures that are not locally finite,
$\sigma$-additivity issues and so on might start with
\citep{Feldman93a} where test spaces (under the name of ``manuals'')
and orthoalgebras are discussed; they are included in more 
general discussions of  convex effect algebras 
in \citet{Bugajski2000a} and of sequential effect algebras 
in \citet{Gudder2000a}.

\subsection{Weak effect algebras from probabilistic equivalence}

Now, I will relate this abstract structure to phenomenological theories, 
by showing that 
one can derive a natural weak effect algebra from any phenomenological
theory (restricted for simplicity to ``locally finite''
phenomenological theories (for which 
each measurement has a finite outcome-set)), and 
mentioning how I think one may  naturally extend any
such weak effect algebra to an effect algebra proper.  

The operation $\oplus$ of the weak effect algebra will be the image,
under our construction, of the binary relations ``OR'' (``$\OR$'') in 
the standard propositional logics (one for each measurement)
of propositions about the outcomes 
of a given measurement.   (This is one justification for calling effect
algebras ``logics''.) 

\subsubsection{Boolean algebras}

In order to describe this construction, we first review Boolean algebras.
A Boolean algebra is an orthocomplemented distributive
lattice.  A lattice is a structure $\langle L, \vee, \wedge \rangle$,
where $L$ is a set, $\vee$,$\wedge$ total
binary operations on $L$ with the
following properties.
Both operations are associative, commutative, and idempotent
(idempotent means, e.g., $(a \wedge a = a)$).  In
addition, together they are {\em absorptive}:
\beqa
a \wedge (a \vee b) = a\;, \nonumber \\
a \vee (a \wedge b) = a\;.
\eeqa
$\vee$ is usually called {\em join}, $\wedge$ is usually called
{\em meet}.  

These properties are satisfied by letting $L$ be any powerset (the 
set of subsets of a given set), and the operations $\vee, \wedge$ 
correspond to $\cup, \cap$.  
For $L= 2^X$ (the power set of $X$) we call this lattice the 
{\em subset lattice} of $X$.
The shape similarity of these sets
of connectives, and the fact that the everyday meanings of 
``join'' and ``meet'' are similar
in meaning to ``union'' and ``intersection'', respectively,
provide a useful mnemonic. 

An important alternative characterization of a lattice is as a 
set partially ordered by a relation we will call $\le$.  
If every pair $(x,y)$ of elements have both a greatest lower bound
(inf) and a least upper bound (sup) according to this ordering, 
we call these
$x \wedge y$ and $x \vee y$, respectively, and the set is a lattice
with respect to these operations.  Also, for any lattice as defined
above, we may define a partial ordering $\le$ such that $\wedge$,
$\vee$ are inf, sup, respectively, in the ordering.  So the two
characterizations are equivalent.

A lattice is said to be {\em distributive} if meet distributes over 
join:
\beq
a \vee ( b \wedge c) = (a \vee b) \wedge ( a \vee c) \;.
\eeq
(This statement is equivalent to its dual (the statement with
$\wedge \leftrightarrow \vee$).)

If $L$ contains top and bottom elements with respect to $\le$, we
call them $1$ and $0$.  They may be equivalently be defined via
$a = a \wedge 1,$ $a = a \vee 0$ for all $a \in L$.

We define $b$ to be a {\em complement} of $a$ if $a \wedge b = 0$
and $a \vee b$ = 1.  

Complements are unique in distributive lattices, not necessarily so
in more general lattices.  When all complements are unique, we write
complementation as a unary relation (operation) $'$;  this relation is
not necessarily total even in distributive lattices with $0,1$.

A {\em Boolean lattice}, or Boolean algebra, is a distributive lattice
with $0,1$, in which every element has a complement.  

Any subset lattice $L=2^X$ is a Boolean algebra, with $0 = \emptyset$
and $1 = X$.  
An important theorem of Marshall Stone says that any
{\em complete, atomic} Boolean algebra may be represented 
as a sub-Boolean algebra of a unique (up to isomorphism) 
subset lattice with a topological structure
(now called the Stone space of the lattice).

\iffalse
Also discuss whether a Boolean algebra can be defined just in terms
of one connective and complementation, with the other connective 
defined.  (I think so!)
\fi

Let us formalize the notion of phenomenological theory as follows.  
\begin{definition}
A
phenomenological theory $\cp$ is a set $\cm$ of disjoint finite sets
$M$, together with a set $\Omega$ of functions (``states'') $\omega$
from (all of) $\union_{M \in \cp} M$ to $[0,1]$ such that for any $M$,
$\sum_{x \in M} \omega(x) =1$.  
\end{definition}
$M$ are the possible measurements;
taking them to be disjoint means we are not allowing any {\em a
priori} identification of outcomes of different measurement
procedures.   Any such identification will be at a higher level of
abstraction, via, for example, identification of probabilistically
equivalent outcomes.  $\Omega$ is the set of phenomenologically
admissible compendia of probabilities for measurement outcomes.
$\cm$) is an
example of what Foulis calls a ``test space'': a set $\ct$ of sets
$T$, where $T$ may be interpreted as operations, (tests, procedures,
whatever you want to call them) and the elements $t \in T$ as outcomes
of these operations.  We will assume 
each $T$ is finite. (Without the interpretation, these are better known
in mathematics as {\em hypergraphs} or {\em set systems}.)  
Call the set of all outcomes $\Lambda := \union T$.
In general test spaces, however,
the $T$ need not be disjoint; here they are. Foulis calls such test
spaces ``semiclassical.''
(Sometimes a weak requirement of {\em
irredundancy}, that no one of these sets is a proper subset of
another, is imposed on test spaces; it is automatic here.)
We may sometimes call this test space
$\cm(\cp)$ (recalling, though, that it does not depend on the
state-set of $\cp$.)  States may be defined on test spaces also, in
the obvious way, as functions $\omega: \Lambda \rightarrow [0,1]$
such that $\sum_{t \in T} \omega(t) = 1$ for any $T$.  
It is only when a phenomenological theory is defined
in such a more general context, where a given outcome may occur in
different measurements, that the question of contextuality (does the
probability of a given outcome depend on the measurement it occurs
in?) arises at the phenomenological level.  By not admitting such a
primitive notion of ``same outcome,'' but distinguishing outcomes
according to the measurements they occur in, the construction we make
will turn out to guarantee noncontextuality of probabilities even at
the later stage where the theory is represented by a more abstract
structure in which the elements (effects, or operations) that play the
role of outcomes may occur in different operations.  We will return to
this point in a discussion of what Gleason-type theorems mean in a
setting such as ours.  But it bears pointing out right now, since the
rest of our discussion ignores it, that the question of whether there
can be convincing reasons for admitting a primitive notion of
``same outcome'' (based perhaps on some existing theory in terms of
which the operations and experiments of our ``phenomenological
theory'' are described) is worth further thought.  A related point is
that test spaces provide a framework in which we can implement a
primitive notion of two outcomes of different measurements being the
same, but we cannot implement a notion of two outcomes of {\em the
same} measurement being the same (up to, say, arbitrary labeling).  
It is not clear why if we can do
one, we should rule out being able to do the other.  A formalism in
which one can do both is that of E-test spaces (the E is for effect).
These are sets, not of sets of outcomes, but of multisets of outcomes.
Multisets are just sets with multiplicity: each element of the
universe is not just in or out of the set, but in the set with a
certain nonnegative integer multiplicity.  In other words, where sets
can be described by functions from the universe $U$
to $\{0,1\}$ (their
characteristic functions), multisets are described by functions from
$U$ to $\N$.  Note that the set of resolutions of unity in an effect
algebra, shorn of its algebraic structure, is an E-test space 
(whence the ``E'' for effect in the name ``E-test space.'').  
Not all E-test spaces
are such that an
effect algebra can be defined on them; those that are are called {\em
algebraic}.  Sufficiently nice E-test spaces are {\em prealgebraic},
and can be completed to be algebraic by adding more multisets
without enlarging the universe (underlying set of outcomes).

To each phenomenological theory we may associate a set of 
Boolean algebras, one for each measurement.  We will call this
set of Boolean algebras 
the ``phenomenological logic,'' or even the
``phenomenologic'' of the phenomenological theory;  note,
though, that it is independent of the state-set  $\Omega$.
These are just the
subset lattices of the sets $M$, or what I previously called the
``propositional logics'' of statements about the results of the measurements. 
We will distinguish them by subscripts on the connectives saying which 
measurement is referred to, e.g. ${\wedge}_M$ (although this is redundant
due to the disjointness of the measurements).

The phenomenological states $\omega$ of $\cp$ 
naturally induce states (which we will also call $\omega$) 
on the logic of $\cp$, 
via $\omega(\{a\}) = \omega(a)$,
$\omega( X) = \sum_{x \in X} \omega(x)$.  They will 
satisfy $\omega(M) = 1$ for each $M$, and $\omega(\emptyset) = 0$.  

We have, for
example ($x$ and $y$ are now subsets of outcomes),
\beq \label{problaw}
\omega(x 
{{\vee}_M} 
y) = \omega(x) + \omega(y) - \omega(x 
{\wedge}_M 
y),
\eeq
(which is equivalent to its dual).

We call the elements of the Boolean algebras of a phenomenologic 
{\em events}, and we will refer to the set of events of $\cp$
as $\cv$.

\begin{definition}
Events $e, f$ are {\em probabilistically
equivalent}, $e \sim f$  in a phenomenological theory if they have the 
same probability under all states:
\beqa
\forall \omega \in \Omega, \omega(e) = \omega(f)\;.
\eeqa
\end{definition}

$\sim$ is obviously an equivalence relation (symmetric, transitive,
and reflexive).  Hence we can divide it out of the set $\cv$,
obtaining a set $\cv/\sim ~=: \ce(\cp)$ of equivalence classes of
events which we will call the {\em effects} of the theory $\cp$.  (We
have dependence on $\cp$, rather than just $\cm$, because although
$\cv$ depends on $\cm$ but not $\Omega$, $\sim$ depends also on
$\Omega$. ) Call the canonical map that takes each element $a \in \cv$
to its equivalence class, ``$e$.''

The images $e(M)$ of the measurements $M$ under $e$ are ``measurements 
of effects.''  Together they form an $E$-test space as defined above
(a set of multisets). 

We now define on this space 
a ``logic'' which is, at least as far as possible, the 
simultaneous 
``image'' under the map $e$ of each of the Boolean algebras $M$,
and show that this logic is a WEA.
To this end,  we introduce a binary operation $\oplus$ on the 
effect space.
\begin{definition} \label{defoplus}
\beq \label{jammin}
e_1 \oplus e_2 := e(a \vee_M b) ,
\eeq
for some $a$ such that $e_1= e(a)$, $b$ such that $e_2 = e(b)$, 
and $M$ such that $a,b \in M$ but $a \cap b = \emptyset$.
\end{definition}
If no such $a,b,M$ exist, 
$\oplus$ is 
undefined on the effect space.
(If they do exist, we will say they {\em witness} the existence
of $e_1 \oplus e_2$.)
As part of the proof of Theorem \ref{theorem: main}
we will show from the definition of the map $e$ via 
probabilistic equivalence and the behavior of probabilities with 
respect to $\vee_M,$ that this definition is independent of 
the choice of $a,b,M$.

Let $\omega^e$ denote
the function from the effects to $[0,1]$ induced in the obvious
way by a state $\omega$ on the Boolean algebra:  effects being
equivalence classes of things having the same value of $\omega$,
we let $\omega^e$ take each equivalence class to $\omega$'s value
on anything  in it.  

One evident property not shared by 
general WEA's is that the set $\Omega(\ce)$ of all states on 
such a WEA is separating:  
\begin{definition}
A set of states $\Omega$ on a WEA $\ce$ is {\em separating} if
for $x,y \in \ce$
\beq
x \ne y \Rightarrow \exists \omega \in \Omega 
(\omega(x) \ne \omega(y))\;.
\eeq
\end{definition}

\begin{theorem} \label{theorem: main}
The set $\ce(\cp)$ of effects of a phenomenological theory $\cp$
with state-set $\Omega$, 
equipped with 
the operation $\oplus$ of Def. \ref{defoplus} and the definition
$1 = e(1_M)$ (for some $M$) constitutes a weak 
effect algebra.  There exist phenomenological theories for which
this is properly weak, i.e. not an effect algebra.  
For all $\omega \in \Omega$
the functions 
$\omega^e$ defined above are states on the resulting weak effect
algebra.  $\Omega^e := \{\omega^e | \omega \in \Omega\}$ is separating on 
$\ce(\cp)$.
\end{theorem}

The proof is a straightforward verification of the axioms and 
the statements about states from 
the definition, and an example for the second sentence.

\noindent
\begin{proof}
We begin by demonstrating 
$\oplus$ is in fact a partial binary
operation.  This is done by verifying the 
independence, asserted above, of the definition
of $\oplus$ from the choice of $a,b,M$ 
and of $1$ from $M$.  Suppose  $e_1 = e(a)= e(c),
e_2 = e(b) = e(d), a,b \in M, c,d \in N, a \ne b,
c \ne d,$ $a \wedge_M b = 0$, $c \wedge_N d = 0$.
Consider any state $\omega$ on 
the set of Boolean algebras which is also in $\Omega$, the
states of our phenomenological theory.  By 
the definition of $e$, 
\beq
\omega(a) = \omega(c) {\rm~~and~~}\omega(b) = \omega(d)\; ;
\eeq
therefore $\omega(a) + \omega(b) = \omega(c) + \omega(d)$.
Now $\omega(a \vee_M b) = \omega(a) + \omega(b)$ because
$a \vee_M b = 0$, and similarly $\omega(c \vee_N d) 
= \omega(c) + \omega(d)$. 
In other words, for any state $\omega \in \Omega$, 
$\omega(a \vee_M b) = \omega(c \vee_N d)$, so
$a \vee_M b$ and $c \vee_N d$ are probabilistically equivalent, and
correspond to the same effect.

Each Boolean algebra contains a distinguished element
$1$;  by the definition of state on $\cp$,
these have probability zero, and one, respectively, in all states.
Hence they each map to a single effect, and these effects we will
call $0$ and $u$ in the effect algebra (verifying later that
$0 = 1'$ in the weak effect algebra, so that it is consistent with
the usual definition of $0$ in a WEA).   Of course,
$\omega^e(1) =1$.  It is also easy to see that 
$\omega^e(x \oplus y) = \omega^e(x) \oplus \omega^e(y)$.
Hence the $\omega^e$ are states, as claimed in the theorem.  
The set $\Omega^e$ is obviously separating.  
To be pedantic, suppose 
there exist effects $x,y$ having $\omega^e(x) = \omega^e(y)$ for
all $\omega^e \in \Omega^e$.  By the definition of $\omega^e$, 
$\omega^e(x)$ is    the common value of
$\omega$ on all $e$-preimages of $x$, and $\omega^e(y)$ is the 
common value of $\omega$ on all $e$-preimages of $y$.  If these 
values are the same for all $\omega^e$, then the preimages of $x$
and of $y$ are all in the same equivalence class, so $x=y$.  
Hence, $\Omega^e$ is separating.

We now verify that $\oplus$ satisfies the weak 
effect algebra axioms.

(EA1) Strong commutativity:  If $a,b \in M$ witness the existence of 
$x \oplus y$ as described in the definition of $\oplus$, 
by symmetry of $\vee_M$ and $\wedge_M$ (which enter symmetrically
in the definition of $\oplus$) 
they also witness the existence of $y \oplus x$
and its equality with $x \oplus y$.  

(WEA2) Weak associativity.
Let $a,b \in M, e(a)=x, e(b) =y, a \cap b = \emptyset$, so that
$a,b$ witness the existence of
$x \oplus y$, and also let 
$c, d \in N \text{ and disjoint, } 
e(c) = z, e(d) = x \oplus y$, so $c,d$ witness the existence
of $(x \oplus y)\oplus z$.  
Similarly let $b',c' \in P$ 
witness the existence of 
$y \oplus z$ and 
$a',f \in Q$  witness
the existence of $x \oplus (y \oplus z)$, so that 
$e(a') = x, e(f) = y \oplus z$, and $a',f$ are disjoint.
Then $\omega^e(x \oplus y) = \omega(a) 
\oplus \omega(b)$ and 
\beq
\omega^e( (x \oplus y)) \oplus z)
= \omega(a) + \omega(b) + \omega(c)\;.
\eeq
Also $\omega^e(y \oplus z) = \omega(b') \oplus \omega(c') = 
\omega(b) \oplus \omega(c)$, 
so 
\beq
\omega^e((x \oplus (y \oplus z)) = \omega(a') \oplus \omega(f)
= \omega(a) \oplus \omega(b) \oplus \omega(c)\;.
\eeq
But $\omega^e((x \oplus y) \oplus z) = \omega^e(x \oplus (y \oplus z))$
for all $\omega^e$ implies $(x \oplus y) \oplus z = 
x \oplus (y \oplus z)$ by the fact that $\Omega^e$ is separating.

(EA3) Define $e'$ to be $e(a')$, for any $a$ such that $e(a) = e$, 
and $a'$ is $a$'s unique complement in the Boolean algebra of 
the measurement $M$ containing it.   
Since for any state, 
$\omega(a')=1-\omega(a)$ and this probability is independent of $a$
as long as $e(a)=e$, $e'$ as thus defined is independent of which 
$a$ is chosen.  Moreover, since $a \wedge_M a' = 0$ 
$e \oplus e' \equiv e(a) \oplus e(a')$ is defined and
equal to $e(a \vee_M a') = e(1_M) = 1,$ so that $'$ as we 
just defined it satisfies (EA3).  

(EA4)  Note that $x \oplus 1$ is equal to $e(a \vee_M 1_M)$, 
for some $M$ containing $a$ and with unit $1_M$, where 
$a \wedge_M 1 = 0$ and  $e(a)=x$.
But each $M$ has a unique $a$ such that $a \wedge_M 1_M = 0_M$, 
namely $0_M$. So an $x$ such that $x \oplus 1$ exists;  it
must be $e(0_M)=0$. 

This proves the first part of the theorem.  
We remark that $1' \equiv e(1') = 
e(0_M)$, so defining $0$ as $e(0_M)$ for
any $M$ coincides with the usual effect algebra definition
as $1'$.  
We now construct 
the counterexample required by the second part.

Consider an empirical theory
consisting of states on the two atomic Boolean algebras:

\beqa \label{example}
\begin{array}{lcccccccccr}
M:& ( & a & & b & )&( & & f &  & ) \\
N:& ( &  & c & & )&(& d ~~~)& ( & g & )
\end{array}
\eeqa

with the indicated $a,...,g$ being atoms of the
Boolean algebras involved (``elementary measurement outcomes''). 
The vertical lining-up of 
parentheses in (\ref{example})
visually indicates conditions we will impose on the theory:
that all states of our phenomenological theory respect 
$\omega(a \vee_M b) = \omega(c)$ and $\omega(f) = \omega(d \vee_N g)$;
further, let our theory contain states with nonzero probability for each of
$a,b,c,d,f,g$.  
There are plenty of perfectly good empirical theories satisfying
these constraints, but $\oplus$ on the effect set of such a theory
will not exhibit strong associativity:
although $e(a) \oplus e(b)$ exists and is equal to $e(c)$, 
and $e(c) \oplus e(d)$ exists and is therefore equal to 
$(e(a) \oplus e(b)) \oplus e(d)$, no effect $h$ exists
with $e(h) = e(b) \oplus e(d)$.
\end{proof}

Although we have shown that the existence of 
witnesses for $x \oplus y$ and $(x \oplus y) \oplus z$ need
not imply it, suppose these hold and also
a witness for $y \oplus z$ exists.  Does this imply
a witness for $x \oplus (y \oplus z)$ exists?  Again, the
answer is no, for the same reason.  Just because there
is something
$h$ 
with $\omega(h) = \omega(b) + \omega(d)$ for all $\omega$,
doesn't mean something probabilistically equivalent to
$h$ is a possible outcome of a measurement which also
has an outcome
probabilistically
equivalent to $a$.  We can illustrate this by adding a 
measurement to the previous example
\beqa
\begin{array}{lcccccccccr}
M:& ( & a & & b & )&( & & f &  & ) \\
N:& ( &  & c & & )&(& d ~~~)& ( & g & ) \\
O:& ( & k & )(  &  & h & &   ~~~)& ( & k & )
\end{array}\;.
\eeqa
Note that the presence of $k$ twice does not indicate
two copies of outcome $k$ (which is not allowed in our
formalism for phenomenological theories' test spaces)
but rather just indicates that $\omega(k) = \omega(a)
+ \omega(g).$   (The diagram also indicates the fact, which 
follows, that $\omega(h) = \omega(b) + \omega(d)$.)
 
The theorem indicates that 
we do not even have the semigroup part of ``partial 
abelian semigroup'', though we have abelianity, 
a weaker associativity (associativity-where-defined),
and some other special properties 
(the
remaining effect-algebra axioms)
that not all PAS's have, 
but all effect algebras do.  

An interesting question is whether the objects thus constructed
share particular properties in addition to those of a weak 
effect algebra.  
We mentioned that the image of
the state-set of the phenomenological theory from which $\ce$
was constructed, which in general could be a proper subset of 
$\Omega(\ce)$, is separating.  As we will see in the discussion of
convex effect algebras, this property has nice consequences.

There is a well-developed and attractive theory of effect 
algebras.  It may therefore seem disappointing that our construction
of a logic on effects yields weak effect algebras, rather than 
effect algebras.  If the following conjecture is true (I suspect
it is straightforward to show) then the theory of effect
algebras may be quite useful in this more general context.

\begin{conjecture}[Completion conjecture for WEA's]
Let $\ce$ be a WEA obtained from a phenomenological theory.
A unique effect algebra $\overline{\ce},$ which we call the
{\rm completion} of $\ce$,  
can be constructed from $\ce$ as follows.  Whenever
one side of the associativity equation exists, adjoin new elements
if necessary to make the other side exist, and impose the equation. 
This can also be characterized as the smallest effect algebra containing
$\ce$ as a sub-weak-effect-algebra (with the latter concept appropriately
defined).
\end{conjecture}

There are some similarities between this conjecture and the fact that
one can make a ``pre-algebraic'' test space into an algebraic one
\cite{Foulis81a}, and similar results involving $E$-test spaces.

The adjunction of these new elements is an interesting theoretical move.
In constructing theories, we often suppose the existence of things that
do not, at least initially, correspond to things in the available 
phenomenology.  The idea of including all Hermitian operators as observables
in quantum mechanics is an example;  as mentioned above, there has been
much discussion of whether all of these actually correspond to anything
observable.  This has motivated the search, often successful, for methods
of measuring various observables that had previously not been measured,
and the development of a general theory, of algorithmic procedures for
measuring certain observables.  The situation with the elements whose
existence is required by weak associativity could prove similar.  Although
the resulting effect algebra is not initially the WEA induced by probabilistic
equivalence from the phenomenological theory, it could motivate the search
for empirical methods of  making measurements which would correspond 
to the additional resolutions of unity which were introduced 
to make the initial WEA into an effect algebra.  In any case, it would be
worth studying the nature of information processing, and information theory
if this is possible, in properly weak effect algebras versus their 
completions.

In the next section, I will discuss the extension of these sorts of
ideas to a framework allowing sequential measurements, and in the
section following that, possible applications.  However, we are now
ready for a few remarks on the significance of Gleason's theorem 
\citep{Gleason57a} in
this context.

\section{Gleason-type theorems in light of probabilistic equivalence}

For quantum systems of finite dimension greater than two, Gleason's
theorem can be interpreted as saying that if mutually exclusive
quantum measurement results are associated with mutually orthogonal
subspaces of a Hilbert space, and exhaustive sets of such measurements
to direct sum decompositions of the entire space into such subspaces,
and if the probability of getting the result associated to a given
subspace in a given measurement is independent of the direct sum
decomposition (measurement) in which it occurs (probabilities are
``noncontextual'') then the probabilities for these results must be
given by the trace of the product of the projector onto the given
subspace with a density operator.

A similar theorem with the resolutions of unity into orthogonal
projectors replaced by resolutions into arbitrary positive operators
has been obtained by \cite{Busch99a}, and independently by Caves,
Fuchs, Renes and Mannes \citep{CFRM2003a,Fuchs2001a, Fuchs2002a}.  
In the next section
we will see how this ``B/CFRM''theorem is a case of a general fact about convex
effect algebras.

Sometimes Gleason-type theorems are used in quantum foundations
arguments to justify the quantum probability laws. Then the question
naturally arises: what justifies the two assumptions that the
probability laws are noncontextual, and that they are associated with
orthogonal decompositions, or positive resolutions of unity, on a
Hilbert space?  Although the construction above via probabilistic
equivalence results in structures much more general than Hilbert space
effect algebras, or effect algebras of projectors on Hilbert spaces,
it automatically results in noncontextual probability laws on the
resulting weak effect algebras.  This argument, of course, starts from
probabilities, so it would be circular to use it to justify
noncontextuality and then turn around and use noncontextuality, via
Gleason's theorem, to ``justify'' the probabilities.  From our point
of view, we have a result that we can fairly elegantly, conveniently
represent {\em any} empirical theory by a set of noncontextual
probability assignments on a certain weak effect algebra (and, if the
completion conjecture is correct, embed this in an effect algebra).
In the case of quantum theory, this general recipe provides {\em both}
the Hilbert space structure {\em and} the trace rule for
probabilities, as a {\em representation} of the compendium of
``empirical'' probabilities (perhaps somewhat idealized by the
assumption that any resolution of unity can be measured) of quantum
theory.

The generalization of Gleason-like theorems to weak effect algebras,
effect algebras, and similar structures are theorems characterizing
the full set of possible states on a given such structure, or class of
such structures.  The example of convex effect algebras will be
discussed in the next section.  In the particular case of a Hilbert
space effect algebra, the import of the B/CFR theorem, from the
operational point of view advanced above, is that the quantum states
constitute the {\em full} state space of the ``empirically derived''
effect algebra.  This observation is perhaps more interesting when one
recalls that in other respects, the category of effect algebras
probably does not have enough structure to capture everything we would
like it to about quantum mechanics: for example, the natural
category-theoretic notion of tensor product of effect algebras
(\cite{Dvurecenskij95a}; see also \cite{Wilce94a,Wilce98a}), applied
to effect algebras of finite dimensional Hilbert spaces, does not give
the effect algebra of the tensor product Hilbert space (or of any
Hilbert space), as one sees from a result in \cite{Fuchs2001a} (for a
similar result involving projectors only, see \cite{Foulis81a}).
Possibly relatedly, a natural category of morphisms for convex effect
algebras, those induced by positive (order-preserving) linear maps on
the underlying ordered linear space (see below), is larger in the
quantum case than the ``completely positive'' maps usually considered
reasonable for quantum dynamics.  Nevertheless for a given Hilbert
space effect algebra, its full state space (space of all possible
states) is precisely the set of possible quantum states.

The role of Gleason-like results depends to some extent on point of
view.  In the project of exploiting the analogies between quantum
states and probabilities, their conceptual role is perhaps more
obvious.  Probabilities are, roughly, ``the right way''
(nonarchimedeanity issues aside) to represent uncertainty, and to
represent rational preferences over uncertain classical alternatives.
In this project, it would be very desirable to see quantum states as
``the right way'' to deal with uncertainty in a nonclassical
situation: the Hilbert space structure perhaps sums up the
``nonclassicality of the situation,'' and the probabilities can be
seen as just the consequence of ``rationality'' in that situation.
For such an interpretation of Gleason's theorem, see for example
\cite{Barnum2000a}
and \cite{CFS2002a}.
This might seem to lay somewhat more stress on formal analogies
between quantum logic and classical logic than the operational
approach I have been outlining;  which constructs the Hilbert space
structure {\em and} the probability state-set simultaneously from
the quantum probabilities.  But after the construction
we can still make if we like the same interpretational
division into ``structure'' (here, Hilbert space effects)
and ``any state on that structure,'' done after the fact,
so that quantum mechanics (or whatever weak effect-algebra theory 
we get) just looks like ``any way to be 
rational in a certain situation,'' though the structure
of the situation is ultimately empirically derived from 
probabilities as well.  
Interestingly, another point of view from
which the Gleason's theorem derivation of quantum probabilities 
takes on a special significance is the relative state interpretation.
Here, one also takes the Hilbert space structure as given, and there
is an implicit assumption that orthogonal subspaces correspond to 
mutually exclusive statements about the system.  Various 
arguments have then been given by various authors, some based on additional 
assumptions, that the
only consistent assignment of probabilities is the standard quantum one.
This suggests that the ``structure of the nonclassical situation'' mentioned
above might be described in terms of measurement outcomes (sometimes
called ``propositions'' or ``properties'') having probability zero or one
(at least in a framework where the outcomes are discrete);  then Gleason's
theorem or analogues for other ``property'' structures, might give the
set of possible probability assignments for such a structure.  This idea
need not be associated with the relative states interpretation.  Indeed,
this is my take on the ``Geneva'' approach to empirical theories 
(rooted in the work of Jauch and Piron on ``property lattices'').

\subsection{Convex effect algebras}
In an ``operational'' view of theories, it is natural to take the
space of operations one may perform as convex.  This represents
mathematically the idea that given any two operations $M_1$ and $M_2$,
we can perform the operation $(\lambda_1 M_1, \lambda_2 M_2)$ (where
$\lambda_i \ge 0, \lambda_1 + \lambda_2 = 1$) in which we perform one
of $M_1$ or $M_2$, conditional on the outcome of flipping a suitably
weighted coin (or, in more Bayesian terms, arrange to believe
that these will be
performed conditional on mutually exclusive events, to which we assign
probabilities $\lambda_1, \lambda_2$, that we believe to be
independent of the system under investigation).  A worthwhile project
is to trace the implications of this assumption, or more general
assumptions involving ways of combining empirical event logic and
probability which might be made at the level of phenomenological
theory, through the probabilistic equivalence derivation of weak
effect algebras, arriving, one hopes, at a notion of convex weak
effect algebra.  We will not work this out here, but rather observe
that one might expect that if we looked at the coin face and saw the
index ``$i$'' and obtained the outcome $a$ of $M_i$, this should
correspond to an outcome $\lambda a$ of $(\lambda_1 M_1, \lambda_2
M_2)$, and that any state should satisfy $\omega(\lambda a) = \lambda
\omega(a)$.

Similar definitions may be made at the level of effect algebra or weak
effect algebra (see below).  
For effect algebras
constructed via probabilistic equivalence from such convex
phenomenological theories, they are consequences of  appropriately
formalized 
phenomenological assumptions of convexity (details will appear elsewhere).  
It would be quite reasonable, also, to
pursue the consequences of imposing a generalized convexity based on a
more refined notion of ``vector probabilities'', or other
representations of uncertainty by nonarchimedean order structures.
Such notions of generalized subjective probabilities (and
correspondingly generalized notions of subjective utility functions)
can result from Savage-like representation theorems for preferences
satisfying ``rationality'' axioms but not certain technical axioms
such as ``archimedeanity'' assumptions that make possible real-valued
representations \citep{LaValle92a,LaValle96a,Fishburn98a}.  Such
``probabilities'' can do things like making statements about the
relative probability of probability-zero events (which events even
standard probability theory on continuous sample spaces cannot always
contrive to consider impossible), conditioning on such events, and so
forth.  In such a construction, the states of the phenomenological
theory would have to take these generalized probabilities as values,
and ``probabilistic'' equivalence would have to be defined in terms of
them.  We will avoid such complications as much as possible here, but
knowing about them may help one understand the role of some technical
conditions in results to be discussed below.

The notion of convex effect algebra has been explored by Gudder and
collaborators.  
\begin{definition}
A {\em convex effect algebra} is an effect algebra $\langle E, u,
\oplus \rangle$ with the additional assumptions that for every $a \in
E$ and $\alpha \in [0,1] \subset \R$ there exists an element of E,
call it $\alpha a$, such that (for arbitrary $\alpha, beta \in 
[0,1])$\\ (C1) $\alpha(\beta a) = (\alpha
\beta) a$, \\ (C2) If $\alpha + \beta \le 1$ then $\alpha a \oplus
\beta a$ exists and is equal to $(\alpha + \beta) a$, \\ (C3) $\alpha
(a \oplus b) = \alpha a \oplus \alpha b$ (again, the latter exists),
\\ (C4) $1a=a$.
\end{definition}
The mapping $a \mapsto \alpha a$ from $[0,1] \times E$ to $E$ is
called the {\em convex structure} of the convex effect algebra.

\citet{Gudder98b} 
showed that ``any convex effect algebra
admits a representation as an initial interval of an ordered
linear space,'' and in addition if the set of states on the
algebra is separating, the interval is generating.  
To understand this result, we review the mathematical
notion of a ``regular'' positive cone (which we will just call cone);
it is basic in quantum information science, for example because the
set of quantum states are the normalized members of a cone; so are the
separable states of a multipartite quantum system, the (unnormalized) 
completely positive maps, the positive maps.  It is also
equivalent to the notion of ordered linear space, as we shall indicate.  

%storage
%end storage

\begin{definition}
A {\em positive cone} is
a subset $K$ of a real vector space $V$ closed under multiplication by
positive scalars.
It is called {\em regular} if it is
\begin{itemize}
\item
Convex (equivalently,
closed under addition:  $K + K = K$.) 
\item
Generating ($K-K=V$, equivalently $K$
linearly generates $V$.) 
\item
Pointed ($K \cap -K = \emptyset$, so that it
contains no nonnull subspace of $V$), \text{ and}
\item
Topologically closed (in the
Euclidean metric topology, for finite dimension).  
\end{itemize}
\end{definition}

(In this definition, we adhere to the common mathematical convention
that when sets are inserted into operations in place of operands, 
the expression refers to the set of things obtainable by inserting
all combinations of elements of the sets into the operations: thus
$K-L$ is $\{k-l: k \in K, l \in L\}$ and so forth.)

Such a positive cone induces a {\em partial order}
$\ge$ on $V$, defined by $x \ge_K y := x - y \in K$. 
$(V,\ge_K)$,
or sometimes $(V,K)$,  is
called an {\em ordered linear space}.  The Hermitian operators on a
finite-dimensional complex vector space, with the positive semidefinite
ordering induced by the cone of positive semidefinite operators, are
an example.  (A relation $R$ is defined to be a partial order
if it is reflexive ($x R x$), 
transitive ($x R y ~\&~ y R z \Rightarrow x R z$) and antisymmetric
($(x R y ~\&~ y R x) \Rightarrow x=y$.)  The partial orders induced
by cones have the property that they are ``affine-compatible'':
inequalities can be added, and multiplied by positive scalars.  If
one removes the requirement that the cones be generating, cones
are in one-to-one correspondence with affine-compatible partial orderings.
In fact, the categories of real vector spaces with distinguished
cones, and partially ordered linear spaces, are equivalent.
\iffalse
(If in addition, the pointedness requirement is removed, the resulting
relation may fail to be antisymetric---it is a preorder, rather than 
an order.)  
\fi

We pause to give motivations, some mathematical and some involving
applications to empirical theories, of some of the seemingly technical
conditions of regularity.  Such a cone may represent, for example, the
set of unnormalized probability states of a system, or a set of
expectation values of observables.  The normalized states may be
generated by intersecting it with an affine plane not containing the
origin.  Convexity is fairly clearly motivated by operational
considerations, such as those in the definition of convex effect
algebra above, or in the desire to have a normalized state set given
by intersecting the cone with an affine plane be convex.  Topological
closure is required so that the cone has extreme rays, and the convex
sets we derive by, for instance, intersecting it with an affine plane,
will have extreme points if that intersection is compact;  then the
Krein-Milman theorem states that these extreme points convexly generate
the set.  (An affine plane is
just a translation of a $d-1$-dimensional subspace: for $d=3$, a plane
in the sense of high school geometry.)  In ``empirically motivated''
settings such as ours, (in which the metric on the vector space will
be related, via probabilities, to distinguishability of states or
operations), one can argue that limit points can be as
indistinguishable as you want from things already in the cone, so
closing a cone cannot have empirically observable effects, and may as
well be done if it is mathematically convenient.  In the presence of
some of the other assumptions, pointedness ensures that the
intersection with an affine plane can be compact.  Its appearance in
the representation theorem for convex effect algebras (presumably
essentially because the convex sets one gets via states tend to be
compact intersections of an affine plane with such a cone) is one
``operational'' justification for pointedness. Pointedness also has a
clear geometric interpretation: if $K \cap -K$ is a vector subspace
other than $0$, then instead of a ``point'' at zero the cone could
have an ``edge'' (if it is a $1$-d subspace), which is why nonpointed
cones are often referred to as ``wedges''; of course the ``edge'' $K
\cap -K$, which must be a subspace, could have dimension higher than
one, in general.  The property of being generating is usually
appropriate because any non-generating cone is generating for a
subspace, and we may as well not drag around the extra dimensions.
When several cones are considered at once, this might no longer be
appropriate.

An {\em initial interval} in such a space is an interval 
$[0, u]$ defined as the set of things between zero and $u$ in 
the partial ordering $\ge_K$, i.e. \
$\{ x \in V: 0 \le_K x \le_K u\}$. It is generating if it
linearly generates $V$ (i.e. anything in $V$ can be written as
a linear combination of things in $[0,u]$).  
It can be viewed as a convex effect algebra, called a
{\em linear effect algebra},  by letting $\oplus$ be 
vector space addition restricted to $[0,u]$ and the convex structure
be the restriction of scalar multiplication.

The representation theorem is to be interpreted as saying that any
convex
effect algebra is isomorphic (as a convex effect algebra) to some
such linear convex effect algebra.
In the case of
finite-dimensional 
quantum mechanics the vector space and cone are $H_{d}$ and
the positive semidefinite cone mentioned above, and
the interval referred to in the representation
theorem is of course $[0, I]$.

In addition to the requirements for states on an effect algebra,
states on a convex effect algebra must satisfy $\omega(\lambda a)
= \lambda \omega(a)$.  

The set of all possible states on a convex effect algebra may 
therefore be characterized via a version of 
Lemma 3.3 of 
%Gudder, Pulmannov{\'a}, Bugajski and Beltrametti
\cite*{Gudder99b}, which describes it for linear
effect algebras $[0,u]$.  First, some definitions.
The dual vector space $V^*$ for real $V$
is the space of linear functions (``functionals'') 
from $V$ to $\R$;  the dual cone $K^*$ (it is a cone in $V^*$) 
is the set of linear functionals which are nonnegative on $K$.
Then $\Omega([0,u])$, the set of all states on $[0,u]$ when 
the latter is viewed as a convex effect algebra, is precisely
the restriction to $[0,u]$ of the set of linear functionals 
$f$ positive on $K$ and with $f(u)=1$ (``normalized'' linear
functionals).    Note that the restriction map is a bijection.
Thus, viewing things geometrically,  the states (restricted functionals)
are in fact in one-to-one correspondence with
the (unrestricted) 
functionals in the intersection of $K^*$ with the
affine plane in $V^*$ given by $f(u)=1$.
Since 
any linear functional on the $d^2$ dimensional vector
space $H_{d}$ of Hermitian operators on $\C^d$
has the form 
$X \mapsto \tr A X$ for some $A$,
while the dual to the positive semidefinite cone in 
$H_{d}$ is the set of
such functionals for which $A \ge 0$
(i.e., the positive semidefinite cone is self-dual ($K=K^*$))
this Lemma tells us that 
the states of 
a finite-dimensional Hilbert space effect algebra are precisely those
obtainable by tracing with density matrices $\rho$; in other words, 
the B/CFRM theorem for POVMs is a case of this general
characterization of states on convex effect algebras.  This
illustrates
the power and appropriateness of this approach (and probably other convex 
approaches, in which similar characterizations probably exist) 
to empirical theories, and to problems in quantum 
foundations.  Gleason's theorem itself cannot be established in this
way, because the effect algebra (which is also an orthoalgebra) of 
projectors is not convex.  However, there may be a natural 
notion of ``convexification'' of effect algebras 
according to which $[0,I]$ is the convexification
of the effect algebra of projectors.  Interesting questions are then,
which effect algebras can be convexified, and for 
which of those (as for the effect algebra of quantum projectors)
convexification does not shrink the state-space.  
Conversely, we might ask for ways of identifying special subalgebras 
of effect algebras, composed of effects having special properties 
like ``sharpness'', perhaps having additional structure such as that
of an orthoalgebra.

\subsection{Sequential operations}
\label{sec: operation algebras}

The operational approach I am advocating suggests that one consider
what general kinds of ``resources'' are available for
performing operations.  Provided both system and observer are
sufficiently ``small'' portions of the universe, it may be reasonable
to suppose that the observer may use yet other subsystems (distinct
from both observer and system) as an ``apparatus'' or ``ancilla'' to
aid in the performance of these operations, that the apparatus may be
initially independent of the system and observer, and that the
combination of apparatus and system may be viewed as a system of the
same general kind as the original system, subject to the same sort of
empirical operational theory, with a structure, and a state, subject
to certain consistency conditions with that of the original system.
(Convexity is a case of this, the ancilla functioning as ``dice.'')
It may be that in some limits some of these assumptions break down,
but it is still worth investigating their consequences for several
reasons: so that we can recognize breakdowns more easily, so that we
may even acquire a theoretical understanding of when and why to expect
such breakdowns, and because we may gain a better understanding 
of why empirical theories valid in certain limits (say, 
small observer, 
small apparatus, small system) have the kind of structure they do.

One such structure, already mentioned, is convexity: provided one can
imagine the observer conditioning the performance of various
operations on events (say, ``dice throws'') independent of the system,
and provided there is a sufficiently rich supply of such events with
different (say, subjective) probabilities that it is a reasonable
idealization to assume one can get essentially any desired convex
combination, then we should include all such operations involving
conditioning on such ``dice'' among our possibilities.
Other such elementary combinations and conditionings of operations
should probably be allowed:  essentially, the set of operations should
be extended to allow including them as subroutines
in a classical randomized computation.  (Of course, this will not
always be appropriate;  for example, it would clearly not be in constructing
examples of theories, that are not even classically computationally universal.)
Among other things, this
might get us the $\oplus$ operation previously obtained as the image
of OR($\vee$) in Boolean propositional logics about each operation's outcomes,
``for free,'' as we can use classical circuitry to construct procedures
whose outcomes naturally correspond to propositional combinations
of the outcomes of other procedures, 
and will have the same probabilities as those propositional combinations.

This leads us to the consider the possibility that the set of 
possible operations be closed under {\em conditional composition}.
This means that given any operation $M$, and set of operations
$M_\alpha$, $\alpha \in M$, there is an operation consisting of
performing $M$, and, conditional on getting outcome $\alpha$ of
$M$, then proceeding to perform $M_\alpha$.  
This assumption is
extremely natural, but nevertheless substantive, in the sense
that one could imagine physical theories that did not satisfy it.
Some outcomes of procedures might destroy the system, or so alter
it that we can no longer perform on it all the procedures we could
before.  Nevertheless, it is worth investigating the structure
of theories satisying the assumption (which certainly applies to the
idealization of finite-dimensional quantum mechanics in terms
of completely positive maps on a given Hilbert space).  The 
structures obtained when conditional composition is not universally
possible might turn out to be understandable as partial versions
of those we obtain when it is always possible, or in some other
way be easier to understand once the case of total conditional 
composability is understood.
An operation in this framework, then, can be viewed as a
tree with a single root node on top, 
each node of which is labelled by an operation and the 
branches below it labelled by the outcomes of the operation,
except that the leaves are unlabelled (or redundantly labelled
by the labels of the branches above them).  The interpretation
is that the root node is the first operation performed, and
the labels of the daughters of a node indicate the operation 
to be performed conditional on having just obtained the 
outcome which labels the branch leading to that daughter. 

From now on, we mean by phenomenological theory a phenomenological
{\em operational} theory closed under conditional composition.
If we extend a
phenomenological theory by making this requirement, then the new
outcome-set contains all finite strings of elements of the old outcome
set.  Given closure under conditional composition, a given string can
now appear in more than one measurement.  In order that the
construction of dividing out operational probabilistic
equivalence can work,
we will have to require that the empirical probability of the string
be noncontextual.  We will also use a different notion of
probabilistic equivalence: $x \sim y$ iff for any $a,b$, $\omega(a x
b) = \omega(a y b)$, where $x,y,a,b$ are all strings.  In our context
the noncontextuality assumption can actually be derived from
the disjointness of ``elementary''
operations (those not constructed via composition) and the assumption
that the choice of operation at node $n$ 
of the tree describing an operation constructed via conditional 
composition cannot affect the probabilities of outcomes corresponding
to paths through the tree not containing node $n$.  This is how one 
might formalize a  generalization of the ``no Everett phone'' requirement
suggested in Polchinski's article:  the probability of an outcome
sequence cannot depend on what operation we {\em would have done} if
some outcome in this sequence had not occurred.   
\iffalse
The assumption can be
justified by assigning each node in the tree describing an operation a
degree, equal to its distance from the root, and giving the
interpretation that a node of higher degree than another corresponds
to an operation performed later in time, so that it is reasonable to
require that the probability of the lower-degree operation's outcomes
is independent of which outcome was obtained in the higher-degree
(later) measurement.  It could even be that the requirement to 
interpret higher-degree operations as later is unnecessary (except
in the case where it's obviously the only reasonable thing, where
the two operations are on the same path from the root).  In this 
case, we would be assuming  
\fi

With suitable additional formalization of the notion of phenomenological
operational theory, and appropriate definitions of $\oplus$ and a 
sequential product on the resulting equivalence classes, one can prove
that dividing probabilistic equivalence out of such a
set of empirical operations, in a manner similar to the
construction of weak effect algebras via probabilistic equivalence,
gives what I will call a {\em weak operation algebra}.  The details
will be presented elsewhere.  Here
I will exhibit the quantum-mechanics of operations as a case of 
a general structure, an {\em operation algebra} (OA), 
which I view as the analogue, for operations,
of an effect algebra.  The structure will be related to the
notion of sequential effect algebra (SEA) studied by 
\citet*{Gudder2000a}, but differ from it in important
respects.  

Since this structure will be a partial abelian semigroup, with extra
structure involving only the PAS operation $\oplus$, with a product
meant to represent composition of operations, and additional axioms
about how the two interact, we will discuss some more aspects of PASes
(following \cite{Wilce98a}) before defining operation algebras.  The
reader might want to keep in mind the algebra of trace-nonincreasing
completely positive maps (with $\oplus$ as addition of maps and the
product as composition of maps) as an example to help understand this
abstraction.

Recall that a PAS is a set with a strongly commutative and strongly
associative partial binary operation $\oplus$ defined on it.  Define a
zero of a PAS as an element $0$ such that for any $a$, $a \oplus 0 =
a$. (Uniqueness follows.)  If a PAS does not have a zero, we may 
freely adjoin one; we henceforth include its existence as part of
a PAS.  A PAS is {\em cancellative} if $x \oplus y = x \oplus z
\Rightarrow y = z$, {\em positive} if $a \oplus b = 0 \Rightarrow
a,b=0$.  The relation $\le$ on a PAS is defined by $x \le y
\Leftrightarrow \exists z~~ x \oplus z = y$.  Part of Lemma 1.2 of
\cite{Wilce98a} is that in a cancellative, positive PAS $\le$ is a
partial ordering.  In such a PAS, we define $T$ as the set of top
elements of the partial ordering (i.e. $T = \{ t \in \co | a \oplus t
\text{ exists } \Rightarrow a = 0\}$).  In a cancellative PAS we
define $x \ominus y$ as that unique (by cancellativity) $z$, if it
exists, such that $y \oplus z = x$.  Define a {\em chain} in 
a partially ordered set $P$ (such as a
cancellative, positive PAS) as a set $C \subseteq P$
such that $\le$ restricted to $C$ is total.

\begin{definition}
An {\em operation algebra}  is a 
cancellative, positive PAS  equipped with a total
binary operation, the sequential 
product, which we write multiplicatively.  With respect
to the product, the structure is
(OA5) a monoid (the product is associative) 
with (OA6) a unit $1$ (semigroup is sometimes used as 
a synonym for this 
unital monoid structure). The remaining axioms
involve the interaction of this monoid structure with the PAS structure.
\\
(OA7) $0c=c0=0$.\\
(OA8) $(a \oplus b)c = ab \oplus bc$\\
$a(b \oplus c) = ab \oplus ac$ (distributive laws).\\
(OA9) $1 \in T$.\\
(OA10) Every chain in $\co$ has a sup in $\co$.
\end{definition}

Note that the sup mentioned in (OA10) is not necessarily 
in the chain.  The existence of an inf (again,
not necessarily 
in the chain, though) for chains, indeed for sets,
is guaranteed by the assumption 
of existence of a zero.
Our structure is not an effect algebra because we do not assume it is
(as a PAS) unital (i.e., has at least one unit).  
A {\em unit} of a PAS is an element $u$ such that
for any $a$, there is at least one $b$ such that $a \oplus b = u$.  In
a cancellative, positive, unital PAS (equivalently, effect algebra)
there is a unique unit, the sole element of the top-set $T$.
Axiom (OA10) might need strengthening in order to obtain some of 
the results one would like.  Notably, we would like to have a 
representation theorem in which the operations belong to a cone in 
a vector space (and thus belong to an {\em algebra} in one of the usual
mathematical senses, the sense of a vector space with an appropriate product).
Aside from belonging to a cone, the special nature of the convex
set of operations in such a representation theorem would be expressed
by an additional requirement, deriving from (OA10),
which would specialize to the
trace-nonincreasing requirement in the case of the quantum operation
algebra (and generalize the initial interval requirement in the analogous
(Gudder-Pulmannov\'a) representation theorem for effect algebras).   

\iffalse
Another interesting property we might wish to impose on operation
algebras is:  
\\
(OA11)
for any $a \in \co$, 
the set of operations {\em compatible} with $b$, (i.e., such that
$\a \oplus b$ exists), has a unique supremum.
\\  
\fi

We shall now show that quantum mechanics provides an example of this
structure.  
We refer to the set of linear operators on $\C^d$ as
$B(\C^d)$.
\begin{proposition}\label{prop: qmeffect algebra}
The set of trace-nonincreasing completely positive linear maps on
$B(\C^d)$, with the identity map $\ci$
as $1$, the map $M$ defined by $M(X)=0$ for every $X$ as $0$, 
ordinary addition of maps as linear operators, restricted to the
trace-non-increasing interval, as $\oplus$, and composition of maps as
the sequential product, forms an operation algebra.  
Its top-set 
$T$ is the set of trace-preserving maps.
\end{proposition}

\begin{proof}
The commutativity (OA2) and associativity (OA1) of $\oplus$ and the
behavior of $0$ (OA7), and the unital monoid structure (OA5 and 6) are
immediate.  
Cancellativity holds for addition in 
any linear space, so since $\oplus$ is here a
restriction of addition on a linear space of linear maps, it is cancellative
(OA3).  It is positive (OA4) because $A + B = 0 \Rightarrow A,B = 0$ 
for $A,B$ in a pointed cone (such as the cone of completely positive linear
maps).   
(OA8) follows from the distributivity of multiplication of
linear operators over addition of linear operators.  The top-set $T$
is the set of trace-preserving operations, which follows from the
easy observation that if you add any operation besides the zero
operation to a trace-preserving operation, the result is not
trace-nonincreasing.  (OA9) follows since the identity operation is trace
preserving.  (OA10), and the statement about the top-set being the
trace-preserving operations,  
may require a topological sort of argument,
and needs to be worked out.  
\end{proof}

Let us note the interpretation of $\oplus$ and $\ominus$ in terms of
the HK representation of a map $\ca$ in terms of operators $A_i$, so
the map acts as $X \mapsto \sum_i A_i X A_i^\dagger$.  Modulo
irrelevant details of indexing, the HK representation sequence $A_i$
is a multiset $[A]$ of operators $A$ such that $A^\dagger A \le 1$.
$\ca \ominus \cb$ exists if there are HK representations $[A], [B]$
such that $[B]$ is a submultiset of $[A]$.  (Equivalently, there
are standard HK representation sequences $A_i$ and $B_i$ such that
$B_i$ is an initial segment of $A_i$, i.e. $\cb(X) = \sum_i A_i X
A_i^\dagger$ where $i$ ranges over the first $k$ $A_i$.)  Thus it is
obvious that $\ca \ominus \cb$ will not always exist.

\iffalse
\begin{proposition}
The quantum-mechanical operation algebra described in 
Proposition \ref{prop: qmeffect algebra} satisfies (OA11).
\end{proposition}
\begin{proof}
For an operation $a$, defined $a'$ as the map  completely positive 
map $a'$ whose action on operators $X$ is 
$a'(X) = (I - a^*(I))^{1/2} X (I - a^*(I))^{1/2}$.  Let $b$ be another 
supplement of $a$, i.e. $b + a$ is tracepreserving.
Then $a^*(I) + b^*(I) = I = \sum_i A_i^\dagger A_i + \sum_
\fi

We define a {\em weak operation algebra} to satisfy all the above
axioms except that associativity is replaced with weak associativity
(whose statement is the same as in the definition of weak effect 
algebra).  With suitable additional formalization of the notion 
of phenomenological operational theory, and definitions of $\oplus$
and sequential product on the equivalence classes, one can show:
\begin{theorem}
The set of equivalence classes obtained by dividing the 
relation of operational probabilistic equivalence defined above
out of a phenomenological operational theory, has a natural 
weak operation algebra structure.
\end{theorem}
A more formal statement of the theorem, and a proof, will be given elsewhere.

Note that if we have operational limits on conditional composition, as
discussed above, we might accomodate that by modifying the notion of
operation algebra (or WOA) to make the multiplicative monoid structure
partial.  It would then be interesting to investigate the conditions
under which this partial structure is extendible to a total one (as
well as the conditions under which a WOA can be completed to an OA).

We can add a convex structure to on OA with little difficulty.
We just introduce a map of multiplication by scalars in $[0,1]$
(i.e. a map from $[0,1] \times \co \rightarrow \co$) such that
the axioms (C1--C4) of convex effect algebras hold, and also
\beqa
(COA15) (\alpha a) b = \alpha (ab) = a (\alpha b)\;.
\eeqa
Again, we will expect such a structure to emerge from an operational
equivalence argument applied to a suitable notion of convex operational
phenomenological theory.

\section{Dynamics and the combination of subsystems in operational theories}
\label{sec: combination}

The operation algebra approach sketched above implicitly includes a
kind of dynamics, although without explicit introduction of a real
parameter for time.  Some operation algebras are probably extendible
in natural ways to have a continuous semigroup structure related to
their sequential product, that might allow for the introduction of a
notion of time, although in this case conditional composition would
probably be restricted by appropriate scheduling constraints.
However, in the operation algebra sketched for quantum mechanics, the
assumption is that any unitary---indeed, any completely
positive---time evolution can be achieved.  The time taken for the
evolution is neglected, and the temporal element of the interpretation
(which applies more clearly to the phenomenological operational theory
than its image, the operation algebra) is only the primitive one that
when one measurement is done conditional on the result of another, it
is thought of as being done after the result of the first is
obtained.  This is the interpretation of the tree structure
describing conditional composition, and since conditional
composition is also basic to the
motivation of this tree structure, it
seems almost tautological.  A stronger interpretation of the tree
structure would be to assume the entire operation at a node is 
finished before any of the daughter node operations begin, so that 
time is associated with distance, in nodes, 
from the root.  A more substantial notion of time might
be introduced in many different ways by adding structure to the
operation algebra, e.g. by some consistent specification of how long
each evolution takes, or by the assumption that each evolution can be
done in any desired finite amount of time.  The latter is a very
strong assumption.  A realistic consideration of these matters would
involve a much more detailed account of the interactions between
apparatus and system that are actually available.  This is an
important part of the project I propose, but I will not pursue it much
here.  It reminds us, though, of one of the important lessons of QIP for
foundations mentioned in Section \ref{sec: peculiar}: that which
operations are possible may depend on the resources available, and
that the beautiful structures one sometimes encounters as operational
theories may be idealized.  In particular, much of the attempt to
implement QIP involves struggling with the limitations imposed by the
limited nature of the subsystems, and interactions, physics makes
available.  It is probably important to try to incorporate these
limitations within the structures used in the operational approach to
characterize systems.  \cite{Barnum2002a} can be viewed as the
beginning of an attempt to incorporate such restrictions, describable
as a limitation of the resources available for control and observation
in terms of a Lie subalgebra of the full Lie algebra $\fsl(d)$
appropriate to arbitary quantum operations, generalizing the notion of
entanglement where the subalgebra is the ``multilocal'' one that is
the direct product of local $\fsl$'s.  ``The physics'' may be though
to include much more than just Hilbert space: preferred tensor product
structures, symmetries, the whole business of representation theory.
Another approach to involving this in operational theories has been the
inauguration, in works such as \cite{Foulis2000a} and
\cite{Wilce2000a}, of a theory of group actions on empirical
structures such as test spaces, orthoalgebras, and effect algebras.

It is virtually certain that when one 
attempts to coordinate operational theories into an overall picture, 
their idealized nature will have to be taken into account, and indeed
perhaps explained in terms of the overall picture.  
Incidentally, this whole discussion raises the question of whether 
the operation algebra induced via probabilistic equivalence is going
to wind up including operations which have strings of more than
one elementary outcome as outcomes, but which are not conditional 
composition trees.  The suspicion that it might is 
 based on the folowing observations.  
While the tensor
product of effect algebras is ``generated'' in a certain
sense by one-round conditional composition, in either direction, 
of local measurements
(it is the effect algebra of the algebraic closure of the 
$E$-test space formed by one-round LOCC measurements), it
contains
also measurements such that, although their outcomes are of course
all product effects, they are not implementable by one-round LOCC.
(In quantum information theory 
these are the measurements somtimes called 
``separable but not LOCC'' measurements.)
The proposed ``operation algebras'' seem to be related to the 
idea of viewing a system's dynamics via a sequence of moments
in time, say, with the outcome-set for a sequence of measurements
being the Cartesian product of the outcome-sets at each moment,
and then generating an operation algebra.  The two-moment 
version of this seems very similar to the tensor-product construction,
except that in the case of time, only one direction of conditional
composition is allowed.  (The corresponding ``directional tensor
product'' has in fact already been constructed, for test spaces, by 
Foulis and Randall.)  

Operations not arising from conditional composition trees
would likely be difficult to interpret 
operationally.  My guess is that the algebraic structure induced
via an appropriate notion of probabilistic equivalence from an
operational phenomenological theory will turn out to be a 
``weak operation algebra,'' where, as with WEA's, the PAS operation
$\oplus$ is weakly rather than strongly associative.  This might 
mean that, while it can be embedded in an operation algebra by 
including the ``nonphysical'' measurements that are not interpretable
in terms of conditional composition, probably` all the measurements
in the WOA will be so interpretable.  (On the other hand, since
the quantum mechanics of operations is already an operation algebra,
that suggests that this problem is less prevalent than one might 
have worried it would be.  Maybe the unidirectional nature of the 
conditional composition actually helps here.  The chief physical
worry about the algebras derived from unidirectional composition is,
in a combination-of-subsystems interpretation, that there exist states
on the resulting effect algebras which exhibit ``forward influence'':
influence in same direction as the classical communication 
required for composition.  Usually theories have ``too many'' states
because they have ``too few'' measurements.  (In fact, allowing 
backward communication allows one to ``detect'' the bizarreness of
such states, ruling them out as states on the two-way tensor
product.  Although, unlike the case of ``ruling out'' the nonpositive
trace-one Hermitian operators that are still around in the tensor
product, by introducing still more (``entangled'') measurements, 
we don't even need to introduce additional outcomes here.) 

\begin{problem}
Does the directional tensor product of test spaces, E-test spaces,
orthoalgebras, or effect algebras, contain measurements not
implementable via one-way LOCC?
\end{problem}

An important part of the project of combining operational empirical
logic and QIP ideas to investigate whether or not physics can provide
an overarching structure unifying perspectives, is to understand the operations available in an
operational theory in terms of interactions with apparatus and/or
environment.  In particular, if we have a way, such as the tensor
product in quantum mechanics, of describing the combination of
apparatus $A$ and system $S$ as subsystems of a larger system $L$, we
will probably want to require that the evolution induced on $S$ by
doing an operation on the larger system is, under appropriate
circumstances, one of the operations our theory describes as
performable on the smaller system.  ``Appropriate circumstances''
probably means that the apparatus should be initially independent of
the system, which in turn requires that the notion of combination of
subsystems have a way of implementing that requirement.  Such
assumptions bear close scrutiny, though, as they may be just the sort
of thing that becomes impossible in certain limits. Some, for example
O'Connell and Lewis \cite{FOL2001a}, 
have argued for the physical relevance of
some situations in which open systems are analyzed without the initial
independence assumption.  Independence works well in the case of
completely positive quantum operations, though: indeed, all such
operations can be implemented via a {\em reversible} interaction with
apparatus.  (It is worth occasionally pondering the possible operational
significance, though, of the existence of positive but not
completely positive maps.)

Direct consideration of categories, such as convex operation algebras
and generalizations of these, that describe dynamics is probably the
most promising way to investigate such questions in terms more general
than quantum mechanics.  Possibly the category-theoretic notion of 
tensor product will be defined for these categories.  One could then
examine, for example, whether when applied to construct the tensor
product of two Hilbert operation algebras, it gives the operation algebra
of CP-maps on the tensor product of the Hilbert spaces.  

To define the category-theoretic tensor product in general requires
the notion of a bimorphism.  For ``small'' categories (whose objects
are sets with additional structure, and whose morphisms are 
structure-preserving mappings), we can define a bimorphism of $A,B$
as function $\phi: A \times B \rightarrow T$, where $T$ is another
object in the category, and $\phi$ has the property that for every
$a \in A, \phi_a: B \rightarrow T$ defined via $\phi_a (b) = 
\phi(a,b)$ is a morphism, and similarly with the roles of $A,B$ 
reversed.  In the category of vector spaces, for example, it is just
a bilinear map.  

\begin{definition}
The tensor product $A \otimes B$ is
a pair $(T, \tau)$, where $T$ is another object in the category 
(also often called the tensor product) and 
$\tau: A \times B \rightarrow T$ is a bimorphism, and any bimorphism
from $A \times B$ factors through $T$ in a unique way, and $T$ is
minimal among objects for which such a $\tau$ exists.
\end{definition}

To say $\tau$ factors through $T$ in a unique way is just to say that
for any bimorphism $\beta: A \otimes B \rightarrow V$, there is a
unique $\phi: T \rightarrow V$ such that $\beta$ is $\tau$
followed by $\phi$.  Minimality in a set means not a subobject of any
object in the set.  Probably the uniqueness of the factorization is
therefore redundant.

There is a partial ``operational'' motivation of this construction
when it is applied to categories like effect algebras, operation
algebras, etc...: the existence of $\tau$ and unique factorization
implement the notion that the two structures being combined appear as
potentially ``independent'' subsystems of the larger system, in a
fairly strong sense that one can do any operation (or get any outcome)
on one subsystem while still having available the full panoply of
operations (outcomes) on the other.  This is also probably linked to
influence-freedom: it certainly is in the effect-algebra or test-space
context, and the relation bears investigation in the general
category-theoretic setting (for categories where the notion of state
makes sense).  However, these operational motivations do not impose
the minimality requirement.

For a variety of operational structures one might use to describe
quantum mechanical statics, including test spaces, orthoalgebras, and
effect algebras, the category-theoretic tensor product (including 
the minimality requirement)  5has been
constructed and it turns out not to be the corresponding operational
structure for the tensor product of Hilbert spaces.  This could
indicate that the structure describing statics requires more
specialized axioms, still consistent with quantum mechanics, and then
the tensor product in this new category, call it $\cz$, will come out
right in the Hilbert space case.  It could also be that the difficulty
is the static nature of the categories.  Indeed, the
category-theoretic tensor product of test spaces or effect algebras
includes measurements whose performance would seem to involve
dynamical aspects.  These are measurements describable as the
performance of a measurement $M$ on system $A$, followed by the
performance of a measurement $M_\alpha$, on $B$, where which
measurement $M_\alpha$ is performed is conditional on the outcome
$\alpha$ of the $A$-measurement.  Indeed, the 
the tensor product of effect algebras must contain all product
outcomes, and in fact it can be characterized as the effect algebra
``generated'' by requiring that it contain all the ``1-LOCC'' (local
operations with one round, in either direction, of classical
communication) measurements just described.  Fuchs'
(\citeyear{Fuchs2001a}) ``Gleason-like theorem for product
measurements'' in fact proceeds by doing this construction for the
case of Hilbert effect algebras.  It turns out to be fairly elementary
to show that it can also be characterized as the minimal
``influence-free'' effect algebra containing all product measurements
(i.e. in which we can do all pairs of measurements one on $A$, one on
$B$, with no communication).  Freedom from influence of $B$ on $A$
means that for all states on the object, the probabilities of the
outcomes of an $A$ measurement, performed together with an independent
$B$ measurement, cannot be affected by the choice of measurement on
$B$.  Influence freedom means freedom from influence in both
directions.  Both of these provide strong operational motivations for
the category-theoretic tensor product in this situation.  The two
motivations are closely related.  Each of the characterizations is
easily established starting from the other, and it also turns out that
a similar construction of a ``directed'' product, in which 1-LOCC
operations are allowed in one direction only, rules out ``influence''
in the direction opposite the communication.  These things are also
true, and were in fact first established for, test spaces and
orthoalgebras.

I think it likely that the best way to resolve the problem of
the effect algebra tensor product not giving the quantum mechanical
effect algebra of the tensor product Hilbert spaces is to go
to a dynamical framework like the operation algebras sketched above.
Nevertheless, it is interesting to try to resolve it in a more
static framework, by adding axioms beyond those of, say, convex effect
algebras.  

The difficulty, in the quantum case, is that the tensor product of
orthoalgebras or effect algebras, while it must contain measurements
of effects that are tensor products of Alice and Bob effects, and,
through addition of effects, all {\em separable} effects, does not
contain ``entangled'' Alice-Bob effects.  The separable effects span
the same vector space $B(\C^d \otimes \C^d) \cong H_{d^2}$ of $d^2
\times d^2$ Hermitian matrices (where A,B both have dimension $d$) as
the full set of effects on $\C^d \otimes \C^d$, but they are the
interval $[0,I]$ in the separable cone, not the interval $[0,I]$ in
the positive semidefinite cone.  Consequently the available states,
while they must be linear functionals of the form $A \mapsto \tr AX$
for $d^2 \times d^2$ Hermitian $X$, are the normalized members of the
separable cone's dual, rather than of the positive semidefinite cone's
dual, so $X$ in the functional $A \mapsto \tr AX$ is not necessarily
positive semidefinite.  The separable cone being properly contained in
the positive semidefinite one, its dual properly contains the positive
semidefinite one's dual, so that not only are we restricted to fewer
possible measurements, but their statistics---even those of
independent $A,B$ measurements---can be different from the quantum
ones (although all quantum states are also possible states).  Stated
in more quantum information-theoretic terms: some nonpositive
operators $X$ are nonpositive in ways that only show up as negative
probabilities or nonadditivity when we consider entangled
measurements: since in the effect-algebra or orthoalgebra tensor
product we don't have entangled measurements available to ``directly
detect'' this nonpositivity, these are admissible states on these
tensor products.  Indeed, as observed in \cite{Wilce92a}, they are
isomorphic to the Choi matrices (block matrices whose blocks $M_{i,j}$ 
are $T(\outerp{i}{j})$) of positive, but not necessarily
completely positive, maps $T$ 
(although the normalization condition (trace-preservation) 
appropriate for such maps is different from the (unit trace)
normalization condition appropriate
for states).  Of course, the nonpositivity of the operator
can be ``indirectly detected'' by tomography using separable effects,
since these effects span the space of Hermitian operators.  

One obvious solution to the problem would be to introduce axioms that
would prohibit this divergence between the existence of entangled
states and nonexistence of entangled measurements.  Mathematically,
this divergence reflects the important fact that the positive
semidefinite versus separable effect algebras on $\C^d \otimes \C^d$
are differentiated by the properties of the corresponding cones: the
former, but not the latter, being self-dual.  Self-duality is a very
natural and powerful mathematical requirement on cones.  However, one
might feel the requirement of self-duality to be too strong and/or not
sufficiently motivated from an operational point of view.  My view is
that self-duality is an important part of the essence of quantum
mechanics, and we should strive hard to understand its operational
motivation.  The cones for classical effect algebras can also be
self-dual: e.g. the algebra of fuzzy sets of $d$ objects
(equivalently, of effects on a $d$-dimensional Hilbert space diagonal
in a fixed basis).  The extent to which self-duality is a
``regularity'' condition or a basic conceptual assumption, and the
meaning of such regularity or conceptual content, for
information-processing tasks, should be understood in both quantum and
classical settings.

An axiom, related to self-duality, violated by the tensor product of
Hilbert effect algebras is the ``purity is testability axiom.''  We
develop some concepts before formulating it.

\begin{definition} An {\em effect-algebra theory} is a pair 
$\langle \ce, \Upsilon \rangle$ where $\ce$ is an effect alebera,
$\Upsilon$ a convex set of states on that effect algebra.
\end{definition}

Here $\Upsilon$ may be smaller than $\Omega(\ce)$, the set
of all possible states on $\ce$. 

\begin{definition}
An effect $t$ {\em passes} a state $\omega$ if $\omega(t) = 1$.  An
effect $t$ is a {\em test} for $\omega$ in theory $\langle \ce,
\upsilon \rangle$ if $t$ passes $\omega \in \Upsilon$ and for no state
$\sigma \ne \omega, \sigma \in \Upsilon$, does $t$ pass $\sigma$.  A
state $\omega \in \Omega$ is {\em testable} in $\langle \ce, \Omega \rangle$
if a test for it exists in $\ce$.
\end{definition}

Let us now assume our effect algebras are convex.

If two tests pass $\omega$, so does any mixture of those 
tests.
Let $t$ be a test for $\omega$, then for $\sigma \ne \omega$,
$(\lambda \omega + (1 - \lambda) \sigma)(t) = 
\lambda \omega(t) + (1 - \lambda) \sigma(t) < 1$, i.e. $t$ cannot
test any mixture of $\omega$ with something else.  

Although we just showed that a test tests a unique state, it is
not necessarily the case that a testable state has a unique test.

Let $t$ test $\omega$;  suppose $\omega = \lambda \sigma
+ (1 - \lambda) \tau$.  Then $1= \omega(t) = \lambda \sigma(t)
+ (1 - \lambda) \tau(t)$.  This implies that $\sigma(t) = \tau(t) = 1$,
hence by the fact that $t$ tests $\omega$, $\sigma = \tau = \omega$.
In other words, only pure (extremal) states can be testable.

We will be interested in 
\begin{axiom} \label{ax: testability}
All pure states are
testable. 
\end{axiom}

To study the consequences of this axiom, we introduce a basic 
notion in convex sets.
\begin{definition}
A {\em face} of a convex set $C$ is an $F \subseteq C$ such that
for every point $p \in F$, all points in terms of which $p$ can
be written as a convex combination are also in $F$.  In other 
words, for $\lambda_i \ge 0, \sum_i \lambda_i = 1$,  
\beq
\sum_i \lambda_i x_i \in F \Rightarrow (\forall i,~ x_i \in F)\;.
\eeq
\end{definition}
Thus a face of $C$ is the intersection of the affine plane it generates with
$C$.  The set of faces, ordered by set inclusion, forms a lattice.
This lattice characterizes the convex set (up to affine isomorphism, 
which is the proper notion of isomorphism for convex sets since affine
transformations $y \mapsto Ay + b$ 
commute with convex combination).

\begin{theorem}
The theory 
$\langle \ce(\C^d) \otimes \ce(\C^d), \Upsilon
\rangle$ violates Axiom 1 unless $\Upsilon$ is contained
in the set of separable states.  In particular, 
$\langle \ce(\C^d) \otimes \ce(\C^d), \Omega(\ce(\C^d) \otimes \ce(\C^d)) \rangle$
violates it.
\end{theorem}
\begin{proof}
We begin by showing that 
the only states testable in $\ce(\C^d) \otimes \ce(\C^d)$
are pure product states. This is straightforward but we give details
anyway.  Let $\tr X = 1$ and 
$\bra{\chi}\bra{\psi} X \ket{\psi} \ket{\chi} \ge 0$ for all
product states $\kett{\phi}{\chi}$, so that 
$A \mapsto \tr AX$ is a state.
Testability means there is a
separable $A$ with trace between zero and one (separable effect)
such that:
\beqa \label{hoedown}
1 = \tr A X \;.
\eeqa
The requirement on $A$ is equivalent to:
\beqa
A = \sum_i \lambda_i \ket{\chi_i} \ket{\psi_i}
\bra{\psi_i} \bra{\chi_i}~ (\lambda_i > 0,
\sum_i \lambda_i \le 1,  
\ket{\chi_i}, \ket{\psi_i} \text{ normalized}).
\eeqa
Thus (\ref{hoedown}) becomes 
$\sum_i \lambda_i 
\bra{\chi_i}\bra{\psi_i} X \ket{\psi_i} \ket{\chi_i} = 1$, which 
can only hold if one of the $\lambda_i =1$,
and for that $i$,  
$\bra{\chi_}\bra{\psi_i} X \ket{\psi_i} \ket{\chi_i}=1$.
Then (dropping the subscript)
\beqa
X = \projj{\chi}{\phi} + X^{\pi,\perp} + X^{\perp,\pi} + X^{\perp,\perp}\;.
\eeqa
This is a resolution of $X$ into components in four subspaces of the
space of operators on $\C^d \otimes C^d$:  the space $\pi,\pi$ of operators on 
the one-dimensional Hilbert space $\pi$ spanned by the pure product
state, the space $\pi,\perp$ of operators taking $\pi$ 
to $\pi^\perp$, the space $\perp, \pi$ going the other way, 
and the space $\perp, \perp$ of operators on $\pi^\perp$.  The middle
two pieces are manifestly traceless, so the last one must be traceless
for $\tr X = 1$ to hold.  However, 
$\tr X^{\perp,\perp} = \sum_{ij}\braa{i}{j} X \kett{j}{i}$ 
in a product basis $\kett{i}{j}$ for $\perp$.
Each $\braa{i}{j} X \kett{j}{i}$
must be positive since $\tr X^{\perp,\perp} A = \tr X A$ for $A \in 
\perp,\perp$.  So for $X^{\perp,\perp}$ to be traceless, they must 
all be zero, and $X = \projj{\chi}{\phi}$ plus possibly some 
traceless stuff which does not affect the induced state.

Thus if Axiom 1 is satisfied, the extremal states of $\Upsilon$
are product states, so $\Upsilon$ is a face of the 
convex set of separable states.
\end{proof}

Note that we {\em can} have a theory on $\ce(\C^d) \otimes \ce(\C^d)$
satisfying the axiom of testability, but {\em only} if the state space is contained
in the dual of the cone generated by the effect algebra.  This suggests
that the axiom, if required of the full state space $\Omega$ of an 
effect algebra, is pushing us towards the idea that the cone be self-dual.
Of course, if the state-set of a theory is smaller that the full set of
all possible states, the testability axiom might be satisfied even for
theories on non-self-dual cones: for example, the theory whose effects are
all the separable ones, and whose states are also all separable.

The axiom of testability is very natural, and it turns out to have a long
history in quantum logic.  In the convex setting, \cite{Mielnik69a} has
certainly used it,  I think Ludwig (\citeyear{Ludwig83a, Ludwig85a}) has too.
Clearly theories which are the full state spaces
of linear effect algebras that are initial intervals 
in self-dual cones satisfy it.
Probably stronger things are necessary
to get self-duality.  

This axiom makes contact with the ``property lattice'' quantum logics
of Jauch and Piron.  See \citet[pp. 220--221]{Valckenborgh2000a} Their
notion of property roughly corresponds to effects (or the analogues in
other quantum structures, since most of their work was done before
effect algebras were formalized in the quantum logic community) $e$
which can have probability one in some states (in our terminology,
effects that pass some states).  Those states are said to ``possess
the property $e$''.  Actually (and relevant to our observation that
tests for a
state are not necessarily unique), they define properties as equivalence
classes of effects that pass the same set of states.
They construct a lattice of properties for an
empirical theory (set of states on some quantum structure).  
\iffalse
During a visit by Piron to David Foulis and Charles Randall in
Amherst, the three worked out the relationship between the
Foulis-Randall version of empirical quantum logic (at that time based
on test spaces and orthoalgebras) and the property structures of Piron
and his school, set forth in the paper \cite{Foulis83a} and well
outlined in \cite{Foulis98a}.  Rather than ``properties'' as just
defined, it is based on ``attributes.''  These are sets $S$ of
outcomes such that there is some state $\omega$ for which $S$ is its
support (the set of outcomes to which it assigns nonzero probability.
However, these are probably closely related to properties as just
defined, although understanding when they coincide may be an open
problem.  
\fi 

Axiom 1 can be viewed as a statement about the relationship of the
lattice of faces of a convex set of states on an effect algebra (i.e.,
the state-set of an effect-algebra theory) to the property lattice of that
theory.

The extremal states are minimal elements of the
face lattice, and the axiom says that there are ``minimal properties''
possessed by those states: minimal in the sense that no other state
posesses them.  I am not certain if this is minimality in the sense
of Piron's property lattice, however, it seems plausible that this
would hold, perhaps under mild conditions.
A generalization of Axiom 1 would assert, for each face of the 
state-set, the existence of a ``property'' of being in that
that face, i.e. an
effect passing precisely the set of states of that face.  
A similar axiom of \cite{Araki80a}
concerns ``filters'' for higher dimensional 
faces, 
but this also involves ``projection postulate-like''
dynamics associated with the filtering.  Araki also uses, as an
assumption, the symmetry or ``reciprocity'' rule, 
satisfied in the quantum-mechanical
case, that can be formulated once a correspondence $\chi 
\leftrightarrow e_\chi$ between
extreme states $\chi$ and tests $e_\chi$ for them has been set up:
\beqa
\chi(e_\phi) = \phi(e_\chi)\;.
\eeqa
It is not clear to me whether the extreme states $\rightarrow$
effects correspondence must be one-to-one
instead of many-to-one
in order to be able to formulate the axiom, or whether one-to-oneness
might be a consequence of it.

Faces play an important role in Ludwig's work as well, as do
statements reminiscent of Axiom 1, so it is quite likely Ludwig's
argument may turn out to be similar.

Araki credits Haag for emphasizing to him the importance of the
reciprocity axiom.  In the second edition of his book,
\cite{Haag96a} includes an informal discussion of the foundations of quantum
mechanics based on the convex cones framework.  He, too, uses
Axiom 1, and a generalization associating faces of the state
space (one-to-one!) with 
``propositions.''  These ``propositions'' are 
effects passing precisely the states
of the face, and minimal among such effects in the sense of a 
probabilistic ordering of effects
\beqa
e_1 \le e_2 := \forall \omega \in \Upsilon ~\omega(e_1) \le \omega(e_2)\;.
\eeqa
This is a different strategy from the Jauch-Piron equivalence class
one for getting uniqueness of the effect associated to a face, but
it is closely related to it.
Jauch and Piron were trying to get by with less reference to probabilities.
Haag also uses the reciprocity axiom, which he argues imposes 
self-duality.  (Haag uses uses the notion of 
self-polarity, but for our type of cone, this is the same as self-duality.  
The polar of a convex body $C$ is the set of linear
functionals $L$ such that $L(x) \le 1$ for all $x \in C$; the polar of
a cone is the negative of the dual cone, since whenever $L(x)$ is 
positive, $L(x')$ is greater than $1$ for $x'$ a large enough positive
multiple of $x$.  Since the negative of a cone is isomorphic to that cone,
a self-polar cone is self-dual.) 

He also gives some operational motivation for an additional assumption,
that of homogeneity of the cone.  This says that the 
automorphism group of the cone acts transitively on its interior.  
(This means that for any pair $x,y$ of interior points, there is 
an automorphism taking $x$ to $y$.)  The 
operational interpretation of elements of the automorphism group is 
presumably as possible conditional dynamics.  
The operational interpretation of 
the assumption of homogeneity, at least for self-dual cones, is probably
to be that any state 
is reachable from any other state by dynamics conditional on
some measurement outcome.  This is not a self-evident requirement, 
but seems natural.  The motivation might be that if you can't prepare any 
state starting from another state, with a nonzero probability
of success, the state space might ``fall apart'' into pieces not reachable
one from the other (orbits of the automorphism group).  
Or maybe while some pieces might still be
reachable from all others, going the other way might not be possible...
the theory would have intrinsically irreversible dynamics, 
even conditionally.  A more detailed study of the potentially bizarre
features of operational theories whose effects are naturally 
represented by a non-homogeneous
cone, or whose state-space generates one, would
be desirable, both with and without the assumption of self-duality.
The ``falling apart'' into orbits of the automorphism group
might be physically acceptable if the theory represented a perspective
involving radical limitations on our ability to prepare states, say:
going from one orbit to another might require a more powerful agent
than the one whose perspective is being considered, but the consequences
of such an agent's actions might be observable by the less powerful
agent.  Quantum entanglement provides an example of such a situation:  the
perspective of the set of local agents, with the power to communicate
classically, allows for pairs of states to exist, such that
they have different statistics for observables implementable by local
actions and classical communication (LOCC), but it is not possible, even
conditional on some measurement outcome, to prepare one starting from
the other via local actions and classical communication.  Of
course, the LOCC perspective of the local agents is not what is 
usually taken as a ``subsystem'' in quantum mechanics, so these sorts
of perspectives can be taken as derivative rather than fundamental;
but perhaps in some types of theories nonhomogeneous perspectives
play a more fundamental role.

In finite dimensions, as Haag points out, homogeneous self-polar cones
are known (e.g. \citep{Vinberg65a}) to be isomorphic to direct products of 
the cones whose faces
are the subspaces of complex, quaternionic, or real Hilbert spaces.
(Extensions of these results to infinite dimensions are obtained in
\cite{Connes74a}.)  The factors in the direct product can be thought of as 
``superselection sectors.;''  classical theory would be recovered 
when the superselection sectors
are all one-dimensional (at least in the complex and 
real cases).   
\cite{Araki80a} obtains a similar theorem except
the effects get represented as elements of a finite dimensional
Jordan algebra factor.  
These are isomorphic to to $n \times n$ 
Hermitian matrices over $\R, \C$, or the quaternions 
$\H$, or a couple of exceptional
cases (spin factors and $3 \times 3$ Hermitian matrices over the
Cayley numbers).  (This is obviously closely related to the cone
representation just described.)
He also gives arguments for picking
the complex case, based on the properties of composition 
of subsystems in the various cases.
Araki's argument is that ``independence'' of the
subsystems should be expressed by $\text{ dim } V = 
(\text{ dim } V_1)(\text{ dim } V_2)$ for the algebras.
But, ``essentially because the tensor product of two skew-Hermitian 
operators is Hermitian'', we have 
$\text{ dim } V >
(\text{ dim } V_1)(\text{ dim } V_2)$ except in trivial cases, when we
take the $V$'s to be the algebras of Hermitian matrices over
real Hilbert spaces $H_1$, $H_2$, and their tensor product.
For $\Q$ there is not even a quaternion-linear tensor product.
The bottom line is that ``the complex field has the most pleasant feature
that the linear span of the state space of the combined system is
a tensor product of [the state spaces of the] individual ones.''

This, too, could use more careful ``operational'' study, but it is
clear there are important operational and probably
information-theoretic distinctions between the cases.
For the real case, the key point is that in contradistinctinon to 
the complex case, states on the ``natural''
real composite system are not determined by the expectation values of
local observables.

We discussed the possible operational motivation or interpretation
for homogeneity,
but said little about self-duality and reciprocity.  I think these 
properties, too, may be related to the ability to coordinate perpsectives
into an overall structure, or the way in which they can be coordinated.
In a ``spin-network'' type of theory, the edges of a graph
with representations of a Lie or quantum group ($\fsu(2)$, 
for spin networks), which are Hilbert spaces.  The vertices are associated
to ``intertwiners'' between those representations.  A state might 
be associated with, say, a partition of the graph by a hypersurface
cutting it into two parts, ``observer'' and ``observed.''  If the hypersurface
has two disconnected parts, the associated Hilbert space will 
involve will be the tensor product of the ones associated with 
the parts;  otherwise, the representation is made out of the representations
labelling the cut edges, in a way determined by the intertwinings at the
vertices between them.  One has
the same Hilbert space whichever piece one takes as ``observer''
vs. ``observed.''  However, it is likely that the role-reversal
between observer and observed corresponds to dualization, and the
result that both correspond to the same Hilbert space will only 
hold in theories in which the structure describing a 
given perspective---here, the Hilbert space associated with the 
surface---is self-dual.  To attempt to actually show something
like this would involve a project of trying to 
construct ``relational'' theories like the Crane-Rovelli-Smolin
theories, but with other empirical theories playing
the role of Hilbert spaces and algebras of observables on them.
A simple first example might be  
``topological classical field theories,'' if these can consistently
be defined.  
In these general ``pluralistic structures'' 
coordinating perspectives, one 
might hope to find a role for self-duality and the reciprocity
axiom, and perhaps homogeneity as well.  For the different empirical
structures associated with  different surfaces to relate to each other
in a ``nice'' way, it might be necessary that the structures be
defined on self-dual cones, or exhibit reciprocity.  Haag says,
``[reciprocity] expresses a symmetry between ``state preparing 
instruments'' and ``analyzing instruments'' and is thus related 
to time-reversal invariance.''  This suggestion, too, bears 
more detailed investigation, perhaps in the same context.
Another relevant point is that quantum groups can roughly be defined
as unital algebras (in the sense of vector space with linear 
associative product, and a unit for the product) with enough additional
structure that their representations have natural notions of dual
representation, and a monoidal product of representations which is
close to being a tensor product, so that it could turn out that 
a broad class of network structures close to the class already considered
by Crane, Rovelli, and Smolin, labelled with quantum group
representations and intertwiners, are actually close to the most general
structures one can build by coordinating operational perspectives. 
This is a line of inquiry that certainly deserves to be pursued further.

While I mentioned above the need to deal with dynamics and system
combination in a structure, such as operation algebra, that is
dynamical from the outset, it is also worthwhile to understand how the
static structure that is available even in dynamical theories (e.g.,
an effect algebra can probably be derived from the same kind of
operational phenomenological theory that can give us an operation
algebra, and the two be closely related) is related to the dynamics.
This suggests that we might ask what kind of ``conditional dynamics''
we can introduce on an effect algebra.  For me, at least, the project
of understanding dynamics in this way is farther along than the more
natural project in terms of operation algebras, so I will discuss it
at more length.

\section{Dynamics on effect algebras: ``Heisenberg'' and ``Schr\"odinger/
 Liouville/von Neumann'' pictures}
\label{sec: dynamics}

We now consider dynamics on effect algebra theories.  A similar
treatment will work for other related structures, e.g. theories on
orthoalgebras, test spaces, or E-test spaces.  Initially we will 
consider ``unconditional dynamics,'' i.e. those preserving total
probability.  These are a special case of ``conditional dynamics,''
in which we condition on the unit of the effect algebra.  We 
refer to resolutions of unity in the effect algebra as 
{\em measurements}.

In a ``Schr\"odinger'' picture (or perhaps we should call it a
``Liouville-von Neumann'' picture), a dynamical evolution on an effect
algebra is represented by an affine mapping $\sigma$, possibly
many-to-one, from states to states.  The affinity of the mapping just
means that it is compatible with the convex structure of the state-set
in the following sense: \beq \sigma( \lambda \omega_1 + (1 - \lambda)
\omega_2) = \lambda \sigma(\omega_1) + (1 - \lambda)
\sigma(\omega_2)\;.  \eeq We will define a {\em Schr\"odinger
dynamics} $\cd$ on an effect algebra to be a semigroup (monoid, with
identity) of such Schr\"odinger evolutions.  These are the ``possible
evolutions'' for a system.  Making the set a semigroup just says that
the composition of two possible evolutions is also a possible
evolution.  Unless explicitly mentioned, we will also assume that a
dynamics is a convex set.  Both of these requirements may be justified
from our operational point of view: but it bears emphasis that this
means thinking of dynamics as the set of evolutions we can, in
principle, make the system undergo.  As an example of a dynamics, take
a finite classical system: its $\cd^*$ contains all the $\sigma_A$
defined by: \beq\label{classical evolution} \sigma_A({\bf p}) = A{\bf
p}\;, \eeq where $A$ is a stochastic matrix.  In finite-dimensional
quantum mechanics, $\cd$ is just the set of (trace-preserving)
completely positive maps on the space of linear operators on the
Hilbert space.  (More precisely it is the set of restrictions of such
maps to the manifold of density matrices; or, even more precisely, it
is the set of induced actions of such maps on the normalized linear
functionals $\omega_\rho$ defined by $\omega_\rho(A) := \tr \rho A$,
since it is these functionals that are the states when quantum
mechanics is viewed in terms of effect algebras.)

We will say that an evolution $\sigma \in \cd$ 
is {\em reversible} if $\sigma^{-1} \in \cd$.  
Thus, the {\em reversible dynamics} $\calr$ of a system, the
subset of $\cd$ consisting of reversible evolutions, is a group and
not just a semigroup.

\begin{proposition}
If $\sigma$ is reversible, then it takes pure states to 
pure states.
\end{proposition}
{\em Proof:}
Let
$\sigma(\psi) = \sum_i \lambda_i \chi_i$.
Then 
$\sigma^{-1} (\sum_i \lambda_i \chi_i) = \psi$.
But by affinity, $\sigma^{-1} (\sum_i \lambda_i \chi_i) = 
\sum_i \lambda_i \sigma^{-1}(\chi_i)$.
By the extremality of $\psi,$ this requires 
$\sigma^{-1}(\chi_i) = \psi$ for
all $i$, whence $\chi_i$ is independent of $i$.  So 
$\sigma(\psi)$ is pure. \QED

(In a quantum-mechanical effect algebra, such evolutions are given
by $\rho \mapsto U \rho U^{-1}$, for unitary or antiunitary
$U$, by Kadison's theorem.  In a finite classical effect algebra
of functions on a set $S=\{1,...,d\}$,
they are given by letting $A$ in (\ref{classical evolution}) be 
a permutation acting on $S$ (so $\sigma$ acts on the functions
(effects) by taking $f$ to the function $f^\sigma$ whose value on 
$x \in S$ is $f(\sigma(x))$).  If we represent the functions by 
diagonal $d \times d$ matrices, the permutations act on these
by conjugation.)

Note that by affinity, any evolution is determined, on the full state
space, by its action on the pure states.

Now consider Heisenberg-like dynamics on effect algebras.  Here we
take (unconditional) 
evolutions to be {\em faithful endomorphisms} of the effect algebra.
These are morphisms $\gamma$ of $\ce$ into itself;  recall from the 
definition of morphism that
$\forall a,b \in \ce, \gamma(a) \oplus \gamma(b) = \gamma(a \oplus
b)\text{ and  } \\ \gamma(1) = 1$.
Reversible evolutions should be
{\em automorphisms} of the effect algebra, that is, endomorphisms that
are bijections.  
(Possibly, we will want to represent dynamics conditional 
on measurement results by not-necessarily-faithful endomorphisms
(i.e. remove the $\gamma(1) = 1$ requirement.)
Moreover we will now want to consider convex effect
algebras; then as with the Schr\"odinger evolutions, all Heisenberg
evolutions are required to be affine, that is, commute
with the process of taking convex combinations.  This is for similar
reasons in both the Schr\"odinger and Heisenberg cases: if the state
preparation procedure (for the Schr\"odinger case) or measurement
procedure (in the Heisenberg case) is a convex combination of two
procedures, this is interpreted as meaning that it involves
conditioning aspects of the procedure on variables which are otherwise
independent from both preparation and measurement and the system
itself.  We should be able to imagine learning the value of the
``dice'' variables either before, or after, evolution, with no
different effect on the resulting conditional states.  
(Thus the assumption
may not be appropriate to apply when such ``external random
variables'' or ``sources of ignorance'' are not available.)

Immediately, natural questions suggests are suggested: for a given
effect algebra, is the set of all possible Heisenberg evolutions
effectively the same as the set of all possible Schr\"odinger
evolutions?  
Or, for a theory $\ct$ 
on $\ce$ whose state-set is not the full $\Omega(\ce)$,
and with a specified Schr\"odinger dynamics, 
are the evolutions in this dynamics all representable as Heisenberg
dynamics?
The same questions could also be asked for reversible
Heisenberg/Schr\"odinger evolutions.  
More precisely, is it the case
that for every outcome $a$, state $\omega \in \cs$, and Schr\"odinger
dynamics $\sigma \in \cd$, there exists a Heisenberg dynamics $\gamma
\in \ch$ such that \beqa \sigma \omega (a) = \omega(\gamma a)\;.
\eeqa Note that the converse is guaranteed when $\ch$, $\cd$ are the
full sets of Heisenberg and Schrodinger dynamics (the maximal sets
satisfying the axioms) on $\ct$; for every $a$, $\omega$, and
Heisenberg $\gamma$, the state $\omega^\gamma$ defined by
$\omega^\gamma(a) = \omega(\gamma a)$ is guaranteed (by the fact that
$\gamma$ is an endomorphism of $\ce$) to satisfy the requirements to
be a state.  However, it is easy to find examples of effect algebras
where the full set of Schr\"odinger dynamics includes some which are
not representable as Heisenberg dynamics.

I will now argue that on the operational conception of measurement and
preparation, 
a very natural assumption about a dynamical theory on an
effect algebra is that only evolutions representable as
Heisenberg evolutions be allowed.  The reason is that from an
operational point of view, we should take as tests any procedures
performed on the system, which yield various outcomes.  This includes
the procedure allowing the system to evolve by one of its allowable
(Schr\"odinger, say) evolutions $\sigma$, and then performing a test
$T$.  This is a procedure we could perform starting {\em before} the
evolution, and each of its outcomes $a$ should also be represented as
an outcome $a^\sigma$ of a test $T^\sigma$, on the unevolved system,
such that \beqa \omega(a^\sigma) = \sigma \omega (a)\; \eeqa If our
effect algebra is required to contain $T^\sigma$, then there is an
endomorphism $\gamma_\sigma$ which takes each $a$ to $a^\sigma$, i.e.,
a Heisenberg representation of the Schr\"odinger dynamics $\sigma$.

This justification could fail if, for example, the effect algebra
corresponding to measurements performable after evolution were
different from the effect algebra for measurements performable before.
Such a situation would require dynamics to be maps from one
effect algebra (or its state space) to another('s).  Another
(generally weaker, in my view) motivation for rejecting the argument
would be a rejection of the operational conception of measurement
taken here.  If measurement, for example, were conceived of as some
magic process taking place instantaneously, then one might not include
the procedure of wait-and-measure in the set of ``primary''
measurements which can be made at time $t$, even if that set of
primary measurements were the same for all $t$.  In either case, the
theory would exhibit a kind of lack of time-translation invariance.
In fact, the assumption that the theory does not lack this weak sort
of time-translation invariance seems closely related to the assumption
behind our introduction of operation algebras: that any operation we
can do now, we can also do conditional on the outcome of another
operation which is done first (in particular, conditional on the
outcome of a top element of the operation algebra, which just means,
following some previous unconditional dynamics).  

A potentially misleading point here is that an endomorphism of the
effect algebra that is not an automorphism maps the full effect
algebra onto a proper sub-effect algebra, so that it might seem to be
``changing the effect algebra'' and therefore violating the assumption
used to justify endomorphisms as dynamics in the first place.
However, the Heisenberg picture is tricky, because it goes in a sense
``backward in time'': it maps the {\em full effect algebra} {\em after
evolution}, with its usual operational interpretation, into the
algebra of effects {\em before} evolution.  Every operation in the
effect algebra is still performable after the evolution; to find out
what the corresponding probabilities after the evolution are, though,
we need to map it into itself, possibly onto a proper subalgebra, and
then evaluate the original state on it.  Whether a subalgebra 
or the full algebra is mapped into the original algebra depends on 
the particular evolution.  By contrast, a Heisenberg-like picture
for a theory in which the operations performable after evolution
were represented by a different effect algebra, say a subalgebra, 
would just map that subalgebra into the original algebra.  {\em No} 
evolution would ever map an algebra isomorphic to the original 
algebra onto itself:  reversible dynamics would be impossible.  

Note that the theory whose state-set is the full state-set of 
$\ce(\C^m) \otimes \ce(\C^n)$, or even the theory
with that (separable) effect-algebra but the states restricted
to the quantum-mechanical ones, have Schr\"odinger dynamics that are
not Heisenberg-representable (e.g. those taking separable states
to entangled states).  This is probably related to their failure to
exhibit other axiomatic criteria (such as the ``purity is testability''
axiom) mentioned above.

As with so many plausible assumptions about operational theories, 
this ``time-translation invariance'' 
is not quite a ``law of thought'':  we could certainly imagine
a lack of time-translational invariance of this sort might 
crop up in physics.  But it is useful to see how it corresponds
with natural formal statements about operational structures.
Moreover, a dynamical theory that does have this sort of invariance
would probably exhibit features with a natural information-theoretic
interpretation. 

One possibility involves distinguishability.  
An important tool in quantum information theory, and QIP theory, 
has been measures of distinguishability of two, possibly mixed,
quantum states.  A copious supply of such measures may be obtained
in a general operational setting using a strategy which has 
proved useful in quantum information theory.  It starts
by considering classical measures of distinguishability of probability
distributions.  
In an effect algebra dynamical theory, 
we simply define the distinguishability
of two states $\rho, \omega \in \cs$ as the maximum over
effect-tests $\Sigma$ (sets of effects $e_i$ such that
$\oplus_i e_i = 1$) of the 
distance between the classical probability distributions
$p_{\Sigma,\rho}$ and $p_{\Sigma,\omega}$ 
induced by $\rho$ and $\omega$
on the outcomes in $\Sigma$:
\beqa
D_{\cf}(\omega, \rho) := \max_{\Sigma} 
D_{cl}(p_{\Sigma, \rho},p_{\Sigma, \omega})\;.
\eeqa 

The question then arises:  
when $D_{cl}$ is nonincreasing
under classical dynamics,  
is the induced $D_{\cf}$ also 
nonincreasing, under the notion of dynamical evolution 
incorporated into $\cf$?
When the dynamics consist of effect-algebra endomorphisms,
the answer would seem to be trivially positive.  (The argument
does not even use the assumption that $D_{cl}$ is contractive.)
For
all the measurements made after evolution map correspond, via the 
Heisenberg-picture endomorphism, to measurements made
before the evolution;  the maximization over measurements
performed after evolution, then, is over a set of 
measurements no larger than that before evolution (smaller,
if the evolution is not an automorphism).   
There has been extensive study of the distance measures
which are contractive under quantum evolutions
(i.e., unital completely positive linear maps 
on operators interpreted as observables, or,
dually, trace-preserving completely positive linear maps
on the state space of such systems).  The contractiveness of
such distances has proven a useful tool, e.g., in establishing
impossibility results in quantum information processing.
For example, the ``no-broadcasting theorem,'' a generalization 
of the no-cloning theorem to mixed quantum states, was first
proved (for nonsingular density matrices) in this way by
\cite{Barnum96a},
although interesting and perhaps more natural
$C^*$-algebraic proofs \cite{Lindblad99a} have now been found.
It may well be that the contractiveness of an appropriate set
of distance measures is a principle that (combined with 
a tensor product structure of system composition which may well,
as it does in the cases of orthoalgebras and effect algebras,
automatically prohibit instantaneous inter-system influence)
render exponential speedup of brute-force search impossible. 

In fact, violation of this sort of ``time-translation invariance''
may lie behind the fact that some versions of ``nonlinear quantum
mechanics'' appear able to speed up NP and \#P problems.  The
combination of a nonlinear evolution law with the usual rules for
quantum measurement ensures that in these theories,
measure-and-then-evolve is not the same as evolve-then-measure; and
indeed, Lloyd and Abrams' algorithms make use of the ability to
increase distinguishability exponentially in time via nonlinear
evolution.  This violation is probably also at the heart of the
theoretical objections to nonlinear quantum mechanics discussed below.
It underlines that the question of how to count ``resources'' and
``information'' become very subtle when such a natural formal
requirement is dropped, and keeping things consistent is difficult.
This is not to say that, like other ``natural'' operational
requirements such as the tensor product law of system combination,
this sort of time-translation invariance can never fail; it might be
an idealization that holds in the limit of perspectives of a certain
type, and it might be necessary to transcend it in certain settings,
perhaps involving cosmology or quantum gravity.  And, our requirement
that all evolutions of the ``starting at time $t$, 
evolve and then measure'' type be included
among the measurements we can make at $t$, may give us a description 
of 
observation that has thrown away information important for
complexity issues (notably, the time it takes to make measurements
via a set of physically easy-to-implement evolutions).
But the discussion
focuses attention on the nature of the problems likely to arise in
making a sensible theory of situations in which it fails.

\section{Tasks and axioms: toward the marriage of quantum information science and operational quantum logic}
\label{sec: applying operational logic}

\iffalse
QIP emphasizes the  relevance of the peculiarities of
quantum mechanics to the performance of tasks.  
QIS's demonstration of the potential worldly power
of these peculiarities is probably closely related to many physical
phenomena, but in a way that is not yet well understood.
\fi

\iffalse Most tantalizing, perhaps, is the potential relevance to
statistical mechanics.  Maxwell's demon can be exorcised within a
classical framework, so although it is this problem that most closely
demonstrates the potential kinship of QM and SM as theories of
perspectives, it is not clear that there is a special quantum
contribution to be made to the problem.  However, in many
situations---such as the harmonic oscillator---quantum mechanics, via
Planck's constant, defines for us a zero of entropy where classically
this is not defined; perhaps this will turn out to be an important
contribution.  (Whether this is likely or not depends on one's view of
those problems for which an infinite-dimensional quantum space is more
convenient: as approximations to an unwieldy (or as yet unavailable)
discrete theory, or as fundamental.)  \fi

QIP emphasizes the usefulness of the conceptual peculiarities of quantum
mechanics to the performance of tasks not classically possible.  This
suggests a strategy of trying to formulate these tasks, or the 
associated concepts, in ways general enough that we might hope to 
characterize different operational theories by whether or not these
tasks can be performed in them, or by the presence or absence of
conceptual phenomena such as:  superposition, complementarity (which may be 
essentially the same thing as superposition), entanglement, 
information-disturbance tradeoffs, restrictions on cloning or 
broadcasting, the nonuniqueness of the expression of states
as convex combinations of extremal quantum states (versus the 
uniqueness classically).  

The impossibility of bit-committment is a candidate for the set of
axioms about of information-processing limitations shared by quantum
and classical mechanics.  This was suggested by Gilles Brassard; Chris
Fuchs and Brassard also speculated that the combination of
no-bit-committment and eavesdropping-proof key distribution might
single out quantum mechanics.  This intriguing suggestion is spun into
a thought-provoking fantasy in \cite{Fuchs2001a}.  In the setting of
$C^*$ algebras, Bub, Clifton, and Halvorson \cite{CBH2003a} have
obtained qualititative confirmation of this hypothesis.  One might
also want to investigate it in more general convex settings, if a good
notion of, for instance, subsystem combination is developed for
classes of these.  Of course, even before the upsurge of interest in quantum 
information science, these conceptual peculiarities were being 
generalized and studied by empirical/operational quantum logic researchers.
For example, superposition has been considered in a test space
setting by \cite*{MKBennett90a};  the nonuniqueness of the extremal decomposition
was studied in a convex sets framework by \cite*{Beltrametti93a}.

Assumptions and tasks involving computation should also be investigated;
I will discuss these in a bit more detail.  In particular, it would
be interesting to establish linkages between complementarity, or superposition,
and computational speedup in some framework more general than quantum and
classical mechanics. 

For information-processing or computation, both dynamical
considerations and composite systems are of the utmost importance.
Since the environment which induces noise in a system or the apparatus
used by an information-processing agent must be considered together
with the system, a notion of composite system is needed.  And notions
of composition are basic to computational complexity, where the
question may be how many bits or qubits are needed, as a function of
the size of an instance of a problem (number of bits needed to write
down an integer to be factored, say) to solve that instance.  Indeed,
the very notion of Turing computability is based on a factorization of
the computer's state space (as a Cartesian product of bits, or of some
higher-arity systems), in terms of which a ``locality'' constraint can
be imposed.  The constraint is, roughly, that only a few of these
subsystems can interact in one ``time-step.''  The analogous quantum
constraint allows only a few qubits to interact at a time.  In general
operational models, some notion of composition of systems, such as a
tensor product, together with a theory describing what dynamics can be
implemented on a subsystem, could allow for circuit or Turing-machine
models involving ``bits'' or other local systems of a nature more
general than quantum or classical systems.  There may be much to learn
from a study of computational complexity in such general systems.

One of the meatiest open problems suggested by QIS, in my view a very
physical though also very abstract problem, is this: why do neither
quantum mechanics nor classical mechanics allow a speedup to
polynomial time of ``brute-force'' search for solutions to problems in
NP.    More precisely, we are asking whether there is something that QM
and the classical description have in common (other than the fact that
they are both embeddable in a quantum description) that prevents them,
given a ``black-box'' which tells us whether a given string is a
solution to a specified instance of a problem or not, and a space of
candidate solutions whose description length is polynomial in the size
of the problem, why no procedure using a quantum or a classical
computer can, for every instance of the problem, find a solution to
that instance using the black-box solution checker a number of times
bounded by a polynomial in the size of the instance.  (This result is
due, for the quantum case, to \cite*{Bennett97b}.)  To make the
general question precise enough for rigorous investigation requires
specifying what such a ``black-box'' would be in the class of
empirical theories (more general than but including quantum and
classical mechanics) in which one is going to ask the question.  It is
clear in the quantum case what such a ``black-box'' should be, but
less so in more general settings.

This is an example of the kind of problem I believe operational
quantum logic has much to contribute to.  One can formulate the
problem in terms of effect-algebras, or weak effect-algebras, with a
model of dynamics, or operation algebras,
or some subcategory of one of these.  Most
likely the set of possible dynamics will be either the full set of
endomorphisms on the effect algebras in question, or a
composition-closed, probably convex 
subset thereof.  To formulate such a problem
conceptually will require implementability of the solution-checking
oracle as a dynamical evolution.  This enables us to formulate a
notion of query complexity in this operational model: an $r$-query
algorithm for computing a function $f$ on a black-box input $x$ ($x$
would be the solution-checking black box, in the special case of
brute-force search for the solution to a problem with polynomical
solution-checking routine) in the model consists of a sequence of $r$
dynamical evolutions chosen from the set of possible evolutions, and
its execution consists of preparing the system in a standard starting
state, applying the first evolution $D_1$ followed by the query
oracle, then $D_2$, then the query oracle, and finishing up, after the
last query, with an arbitrary measurement, interpreting the result via
some fixed (polynomial-time classically computable, say) algorithm as
the value of $f$.  In some models, of course, one can't even compute
in polynomial time---or at all---some polynomially computable
classical functions.  This would restrict the nature of the inputs
$x$, and drastically alter the theory of query computation relative to
the classical case.  In a more tractable situation, holding for
example for quantum computation, all $x$ would be black-box
implementable as circuits in the operational model.  

To go beyond
query computation to explicit algorithms such as Shor's factoring
algorithm (to give a quantum example) requires a notion of
computational resources required to perform a given dynamical
evolution or measurement.  One way of specifying such a notion is by
specifying a set, possibly infinite, of dynamical evolutions to which
we ascribe unit cost, and a specification, say, of a set of
measurements viewed as computationally easy.  More generally, we
might specify a function on dynamical evolutions and 
measurements.  Precisely what we will
want to do here may depend on which of several variant computational
models we want to implement, depending on how we allow ourselves to
interface the given operational model with ``classical'' computation.
Perhaps most generously, we might specify a set of
measurements-with-conditional-dynamics (``instruments'') viewed as
taking unit computational time, and allow the conditioning of further
dynamics and measurement on the results of the measurement in
question.  Subtleties would arise in counting the computational cost
of the classical manipulations required to condition in a specified
way, though probably a satisfactory solution using a standard
classical computational model and counting one elementary operation in
that model as costing the same as one in the general operational
model, would work (at least if the general operational model can
simulate classical computation one-for-one, or at least polynomially).
Most simply, we might just perform the algorithm in the general
operational setting via evolution without explicit measurement and
classical control, and specify a ``standard'' measurement to be
performed at the end (along with a standard procedure, or set of
allowable procedures, for mapping the measurement result to the set of
possible values of the function being computed).  
For explicit algorithms, in non-query models, it is important that
not just any measurement be allowed at the end, since if the dynamics
consists of all effect-algebra endomorphisms, say, 
any computation can be done by making
one measurement.

Using such models of query complexity and/or computational
complexity in some fairly general class of
operational theories, I think we are likely to find intuitively
meaningful, very general properties of operational physical theories,
shared by quantum and classical mechanics but also by a wider class of
theories, which forbid, for conceptually clear reasons,
polynomial-time brute-force search.  These properties may turn out to
be linked to other properties of theories.  Some possibilities are the
second law of thermodynamics (impossibility of a {\em perpetuum
mobile}), or the impossibility of instantaneous signalling between
subsystems of a composite system.  Richard Jozsa has suggested that
the impossibility of speedup of brute-force search (in fact, of
NP-hard problems) to polynomial time could serve as a constraint on
proposed new physics.  In this regard, \cite['s]{Abrams98a}
demonstration that theories of ``nonlinear quantum mechanics'' do
permit such speedups is relevant.  It is particularly interesting in
light of the fact that nonlinear quantum mechanics also appears
inconsistent with the second law of thermodynamics \citep{Peres89a}.
Moreover \cite{Polchinski91a} has argued that a class of nonlinear
modifications of quantum mechanics (which includes Weinberg's
\citeyear{Weinberg89a} proposal) either allow superluminal signalling
(because they allow instantaneous signalling between subsystems, which
may be spacelike-separated) or, if they do not allow superluminal
signalling, allow something he calls an ``Everett phone.''  This
latter phenomenon involves a sequence of spin measurements, in which
the outcome of a later spin-measurement, conditional on the first
spin-measurement having resulted in spin up, depends on what the
observer {\em would have} proceeded to measure if the first
measurement had resulted in spin down.  Polchinski discusses the
Everett phone in the framework of the ``relative states''
interpretation of quantum mechanics.  It would appear to be
inconsistent with a more standard view of quantum mechanics, for
example as an ``operational theory'' of the sort we have been
considering.  On such a view, the probability of a measurement result
is taken to be independent of the context in which it is
measured---i.e., of the other outcomes tested.  (Such
context-independence is automatic on the operational point of view
described above: it is enforced by the definition of effects as
probabilistic equivalence classes of measurement outcomes.)

\section{Conclusion}

If we take a broad view of work on the foundations of quantum mechanics, 
I think we can recognize that
important insights have been achieved by pushing the various points of
view to see how much they can accomplish, how rigorously their
insights can be formulated, what hidden assumptions may lurk.  
Equally important insights are to be had by stepping back
from the project of promoting an individual point of view on quantum
foundations as ``the answer,'' and looking at how the insights derived
from different points of views may relate, acknowledging that none of
us yet has the answer, and the path toward it may require combining
insights from several points of view.  Quantum information science
provides an arena in which such putative insights and their
relationships can be analyzed with appropriate quantitative tools, and
in terms of information-processing concepts and tasks that promise to
have real worldly and/or physical significance.

In this paper, I have promoted a particular project for harnessing the
concepts of quantum information science to the task of illuminating
quantum foundations.  This project is to generalize
tasks and concepts of information science beyond the classical and the
quantum, to abstract and mathematically natural
frameworks that have been developed for representing empirical theories;
and to use these tasks and concepts to develop axioms for such theories,
having intuitively graspable, perhaps even practical, meaning, or
to develop a better understanding for the operational meaning of existing
axioms.  Moreover, I have emphasized a particular strategy for this
project, which begins with 
an ``operational'' approach to describing empirical theories, 
taking the probabilities for various outcomes of operations
one may do on the system as primary, and, via probabilistic 
equivalence, making connections to the more
abstract structures of convex effect algebras 
and convex operation algebras,
which are closely linked to the convex frameworks used to good effects
by the authors of many of the other papers in this volume (Bacciagaluppi,
Busch, and Hardy spring to mind).  
These approaches are particularly relevant to the
project because the structure describing an empirical 
theory depends on the agent
doing the operations whose probabilities it represents:  it is 
``perspectival.''  It is precisely the possibility of
coordination of different
agents' perspectives into an ``objective'' framework, or the impossibility
of doing this in a ``realistic'' way that produces an ``objective'' framework,
that is the key issue in the foundations of quantum mechanics.  So
these operational, perspectival structures, ``operational quantum 
logics'' or whatever you'd like to call them, are appropriate for 
studying these coordination issues.  Since information itself involves
relations between parts of a system, say between an information-gathering
and using agent and another subsystem it is gathering and using information
about, the transmission, preservation, and use of information, and the
network of relations it involves between subsystems, is likely to be
crucial for understanding the nature, and the possibility or impossibility,
of such coordination between perspectives.  That is why the generalization
of tasks and notions concerning information from classical theory to 
quantum theory and beyond to the more general operational structures
discussed here is likely to be useful in this foundational project, 
and why quantum information theory in particular provides a both a model
for that part of the project, and a fertile source of particular axioms,
ideas, and constructions, for use in it.  Conversely,
``integrability of perspectives into a coherent whole,''
is also a possible source of axioms about the nature of perspectives
(self-duality or homogeneity of the cones used to represent them?), 
how they combine (via tensor products or some other rule?), that may 
ultimately help illuminate the significance of the laws of physics, 
and the flow and uses of information in physical systems.

\section*{Acknowledgements}
Discussions over the years with Carlton Caves, Dave Foulis, 
Chris Fuchs, Leonid Gurvits, Lucien Hardy, Richard Jozsa, 
Eric Rains, R\"udiger Schack, and Alex Wilce, 
among others, have influenced
my thoughts on these matters.  Chris Fuchs brought the work of
Bilodeau to my attention.  The epigram from Cormac McCarthy appears
also on Carl Caves' homepage, so that may be where I got it
from although it also leaped out at me when I read the book.

% Bibliographic references with the natbib package:
% Parenthetical: \citep{Bai92} produces (Bailyn 1992).
% Textual: \citet{Bai95} produces Bailyn et al. (1995).
% An affix and part of a reference:
%   \citep[e.g.][Ch. 2]{Bar76}
%   produces (e.g. Barnes et al. 1976, Ch. 2).
\bibliographystyle{elsart-harv.bst}
%\bibliography{found}
%\iffalse
%\begin{thebibliography}{}

% \bibitem[Names(Year)]{label} or \bibitem[Names(Year)Long names]{label}.
% (\harvarditem{Name}{Year}{label} is also supported.)

%\bibitem[]{}

%\fi

\end{document}